\documentclass{emulateapj}

\newcommand{\ugriz}{\mbox{$ugriz$}}
\newcommand{\ug}{\mbox{$u\!-\!g$}}
\newcommand{\gr}{\mbox{$g\!-\!r$}}
\newcommand{\ri}{\mbox{$r\!-\!i$}}
\newcommand{\iz}{\mbox{$i\!-\!z$}}
\newcommand{\photo}{{\tt PHOTO}}
\newcommand{\astrom}{{\tt astrom}}

\slugcomment{}
\shorttitle{SDSS Southern Coadd} 
\shortauthors{Annis et al.}

\begin{document}

\title{The SDSS Coadd: 275 deg$^2$ of Deep SDSS Imaging on Stripe 82}

\author{
James Annis\altaffilmark{1},
Marcelle Soares-Santos\altaffilmark{1},
Robert H. Lupton\altaffilmark{2},
Michael A. Strauss\altaffilmark{2},
Andrew C. Becker\altaffilmark{3},
Scott Dodelson\altaffilmark{1,}\altaffilmark{4},
Xiaohui Fan\altaffilmark{5},
James E. Gunn\altaffilmark{2},
Jiangang Hao \altaffilmark{1},
\v{Z}eljko Ivezi\'{c}\altaffilmark{3},
Sebastian Jester\altaffilmark{1,}\altaffilmark{6}
Linhua Jiang\altaffilmark{5},
David E. Johnston\altaffilmark{1},
Jeffrey M. Kubo\altaffilmark{1},
Hubert Lampeitl\altaffilmark{1,}\altaffilmark{7},
Huan Lin\altaffilmark{1},
Gajus Miknaitis\altaffilmark{1},
Hee-Jong Seo\altaffilmark{8},
Melanie Simet\altaffilmark{4},
Brian Yanny\altaffilmark{1}
}

\altaffiltext{1}{Center for Particle Astrophysics, Fermi National Accelerator Laboratory, P.O. Box 500, Batavia, IL 60510, USA}
\altaffiltext{2}{Department of Astrophysical Sciences, Princeton University, Princeton, NJ 08544, USA}
\altaffiltext{3}{Department of Astronomy, University of Washington, Box 351580, Seattle, WA 98195, USA}
\altaffiltext{4}{Kavli Institute for Cosmological Physics, Chicago, IL USA}
\altaffiltext{5}{Steward Observatory, 933 North Cherry Avenue, Tucson, AZ 85721, USA }
\altaffiltext{6}{currently at the Federal Ministry of Education and Research, 53170 Bonn, Germany}
\altaffiltext{7}{currently at the Institute of Cosmology and Gravitation, University of Portsmouth, Portsmouth, PO1 3FX, UK}
\altaffiltext{8}{Berkeley Center for Cosmological Physics, LBL and Department 
of Physics, Universityof California, Berkeley, CA, USA94720.}

\begin{abstract}
We present details of the construction and
characterization of the coaddition of the Sloan
Digital Sky Survey  Stripe 82 \ugriz\ imaging data. 
This survey consists of $275$ deg$^2$ of
repeated scanning  by the SDSS camera of $2.5\arcdeg$ of $\delta$ over
$-50\arcdeg \le \alpha \le 60\arcdeg$ centered on the Celestial Equator.
Each piece of sky has $\sim 20$ runs contributing and thus
reaches $\sim2$ magnitudes fainter than the
SDSS single pass data, i.e. to $r\sim 23.5$ for galaxies. 
We discuss the image processing of the coaddition, the modeling of the PSF,
the calibration, and the production of standard SDSS catalogs.
The data have $r$-band median seeing of 1.1\arcsec, and are calibrated to 
$\le 1\%$. Star color-color, number counts, and psf size vs modelled size
plots show the modelling of the PSF is good enough for precision 5-band photometry.
Structure in the psf-model vs magnitude plot show minor psf mis-modelling
that leads to a region where stars are being mis-classified as galaxies,
and this is verified using VVDS spectroscopy.
As this is a wide area deep survey there are a variety of uses for the
data, including galactic structure,
photometric redshift computation, cluster finding and
cross wavelength measurements, weak lensing cluster mass calibrations,
and cosmic shear measurements.
\end{abstract}


\section{Introduction}
The Sloan Digital Sky Survey \citep[SDSS;][]{york} saw first light
in 1998 with the goal of obtaining CCD imaging in five broad
bands \ugriz\ over 10,000 deg$^2$ of high-latitude sky in the North 
Galactic Cap, plus spectroscopy of one million galaxies and one 
hundred thousand quasars over this same region. 
In addition, the SDSS imaged a 275 deg$^2$ 
region on the Celestial Equator in the Southern Galactic Cap. 
This region is called ``Stripe 82'' and was imaged multiple times 
during the Fall months when the North Galactic Cap was not observable. 
The SDSS single pass data reach $r\sim22.4$ and has median 
seeing of 1.4\arcsec\ in $r$, but by aligning 
and averaging (``coadding'') the Stripe 82 images 
we reached $\sim$ 2 magnitudes deeper and median seeing 
of $\sim$ 1.1\arcsec.
The intent was to use this deep survey to understand
the single pass data at its limits and to do science at fainter
magnitudes or correspondingly higher redshifts.
Such analyses benefit from our image processing approach, as opposed to 
catalog-level methods, because objects below the detection limit of 
individual single pass images can be detected and measured.
A brief description of the Stripe 82 data and coadd was presented in the SDSS 
Seventh Data Release (DR7) paper (\cite{dr7}; see also, \cite{jiang2008}). 
Here we give a full report, detailing the features in he coaddition 
process.

The SDSS uses a dedicated wide-field 2.5m telescope \citep{theTelescope}
located at the Apache Point Observatory (APO) near Sacramento Peak in
Southern New Mexico. 
The telescope imaging instrument \citep{theCamera} is a wide-field 
camera with 24 $2048 \times 2048$ $0.396\arcsec$ pixel scale CCDs. SDSS
images the sky in drift scan mode with the five filters in the 
order $riuzg$ \citep{fukugita1996}.
Imaging is performed with the telescope tracking great circles 
at the sidereal rate; the effective
exposure time per filter is 54.1 seconds, and 18.75 deg$^2$ are 
imaged per hour in each filter.  The images are mostly
taken under good seeing conditions 
on moonless photometric nights \citep{hogg2001}.
For stellar sources the 50\% completeness limits of the images are
$u,g,r,i,z = 22.5, 23.2, 22.6, 21.9, 20.8$, respectively \citep{dr1},
although these values depend on seeing and sky brightness.  
The image processing pipeline determines the
astrometric calibration \citep{astrom}, then detects objects and measures
their brightnesses, positions and shapes \citep{lupton2001,edr}.  
The astrometry is good to 45 milliarcseconds (mas) rms per coordinate at 
the bright end \citep{dr7}.
The photometry is calibrated to an AB
system \citep{okegunn}, and the zero-points of the system are
known to 1--2\% \citep{dr1,dr2}.  The photometric calibration
is done in two ways, by tying to photometric standard stars 
\citep{smith2002} measured by a separate 0.5m telescope on site 
\citep[the PT telescope;][]{Tucker2006,Ivezic2004}
and by using the overlap between adjacent imaging
runs to tie the photometry of all the imaging observations together, 
in a process called ``ubercalibration'' \citep{ubercal}.  
Ubercalibration  zero-points on each stripe have rms error of
$\sim 2\%$ in $u$ and $\sim 1\%$ in $griz$.

SDSS data is obtained as runs, where a run is a single continuous 
drift scan obtained
on a single night. A survey stripe is one camera width wide, 
about $2.5\arcdeg$.
Two interleaving runs called strips are necessary to complete a stripe
as the camera focal plane is sparsely populated. 
These strips are denoted either
N or S, depending if the telescope boresight is pointed half a CCD width 
north or south of the stripe equator.
A run contains 6 columns of data through the five \ugriz\ filters, and
a single filter data set is called a scanline.
Each scanline is a 13\arcmin\ wide continuous stream of data 
that we  arbitrarily 
chop into overlapping 10\arcmin\
long frames.  A frame is a single image in a single bandpass, and
has a geometry of 1489 rows and 2048 columns, at a pixel scale of 
0.396"/pixel. 
A field is the set of \ugriz\ frames of the same piece of sky, 
disregarding the fact that they were obtained over 8 minutes of time. 
For Stripe 82 in particular, there is a unique mapping of RA and Dec
into survey constructs. The row number of a field corresponds to RA, 
as does the field number. The column number of a field corresponds 
to Dec, as does the camera column number. 
The fields overlap along the RA direction by 124 rows per field due to 
a repackaging of the same pixel data during data acquisition. 
They overlap along the Dec direction by a small amount on either edge 
of the field due to re-observation of sky by 
slightly overlapping scanlines.
The coadd runs are artificial, so we adopted run number 100006  
as the south strip and 200006 as the  north strip arbitrarily. 
These were later renamed as 106 and 206 for convenience.
These runs interleave, but for constant column
number,  run 206 is at a higher Dec than run 106. 
Each run is 800 fields long and goes in increasing field number from 
the West to East, from low to high RA.
For more information on the SDSS nomenclature and 
technical terms, see \cite{edr}.

Stripe 82 is the SDSS stripe along the Celestial Equator in the Southern 
Galactic Cap.  It is 2.5\arcdeg\ wide and covers
$ -50\arcdeg \le RA \le +60\arcdeg$, so its total area is 275 deg$^2$.
Stripe 82 can be observed from APO at low airmass from September 
through November, is accessible from almost all ground-based telescopes 
for subsequent spectroscopic and photometric 
observations and, except near its RA ends, 
has low Galactic extinction \citep{dust}.
Stripe 82 was imaged by the SDSS multiple times in the Fall months 
and through 2004, these data were taken only under
optimal seeing, sky brightness, and photometric conditions (i.e.,
the conditions required for imaging in the main Legacy Survey; 
\cite{york}). There were 84 such runs.
In 2005-2007, 219 additional imaging runs were taken on
Stripe 82 as part of the SDSS supernova survey \citep{sdssSN}, 
designed to discover Type Ia supernovae at $0.1 < z < 0.4$. 
The supernova survey was carried out on most usable nights,
with the exception of the five brightest nights around each full moon.
Therefore, these data were often taken under less optimal conditions: 
poor seeing, bright moonlight, and/or non-photometric skies.  

Both reduced images and catalogs  from all 303 runs 
covering Stripe 82 were made available as part of the
SDSS DR7 \citep{dr7}, in a database called {\tt Stripe 82}. The data 
can be accessed both 
the Data Archive Server (DAS) and the Catalog Archive Server (CAS). 
  We carried out a coaddition of the repeat
  imaging scans, photometric or not,  on Stripe 82 taken through Fall 2005.
Data taken after that date were excluded as for the most part they had not
been taken when we were processing the coadd.
  The coaddition includes a
  total of 123 runs, covering any given piece of the $275$ deg$^2$
  area between 20 and 40 times.
  The S and N strip runs are designated 100006 and 
  200006, respectively, in the DAS, and 106 and 206 in the CAS database. 

We designed the coaddition program so that the output image format
allowed us to run the SDSS standard measurement code, \photo\ 
\citep{lupton2001,edr,photo-lite}, on the coadd images. 
This was important because: a) \photo\ has
algorithms which had been extensively tested by the SDSS 
collaboration over the years; and b) the resulting data products are   
conveniently structured for joint analyses and comparisons with the 
single pass data.
Our method considers the repeat scans of Stripe 82  to be
noisy, distorted realizations of the true sky.
The aim is to make our best estimate of the true
sky as it would have been seen by a perfect SDSS camera on a 
larger telescope. 
Starting with the list of runs on 
Stripe 82 taken from the start of the survey to the Fall 2005 season, 
those fields of reasonable seeing (FWHM), transparency ($T$), 
and sky noise ($\sigma_s$) were
selected for use in the coadd.
The individual runs were remapped onto a uniform
astrometric coordinate system. 
Interpolated pixels (due, e.g., to cosmic rays or bad columns)  
in each individual run were masked and the sky was subtracted 
from each frame.
The images are coadded with weights that depend on FWHM, $T$ 
and $\sigma_s$,
providing optimal signal ratio to noise for point sources.
\photo\ relies on an accurate point
spread function (PSF) model for both stellar and galactic 
photometry \citep{lupton2001}.
Rather than remeasuring the PSF on the coadd
images we computed the PSF by constructing 
  the suitably weighted sum of the PSFs made by \photo\ for each run.
The coadded images were run through \photo\
  yielding the catalog made available in the CAS {\tt Stripe82}
  database.  

When using the coadd data for science it is important, 
just as with the main survey, to use the various 
processing flags associated with each detected object to reject
spurious objects and to select objects with reliable photometry
 (as recommended, for example,  by \cite{richards2002}).
Since the coadd data was run through the SDSS pipelines,
the standard flag set is available for all objects. 
However, some objects at magnitudes $< 15.5$ that are saturated
do not have the saturated flag set, so 
we recommend a magnitude cut to avoid them.

In this paper we first consider the observations (\S\ref{obs}), 
then describe the coadd image (\S\ref{imagecreation}) and 
catalog creation (\S\ref{catalogcreation}). In section \S\ref{results}
we present the results of our quality assurance tests and 
explore the features of this data set, highlighting the improvements
in depth and seeing due to the coaddition process. 
We discuss the applications and science outcomes in \S\ref{science}
and conclude in \S\ref{conclusion}.

\section{Observations}\label{obs}
\begin{figure}[!h]
\centering
\includegraphics[width=1.0\columnwidth]{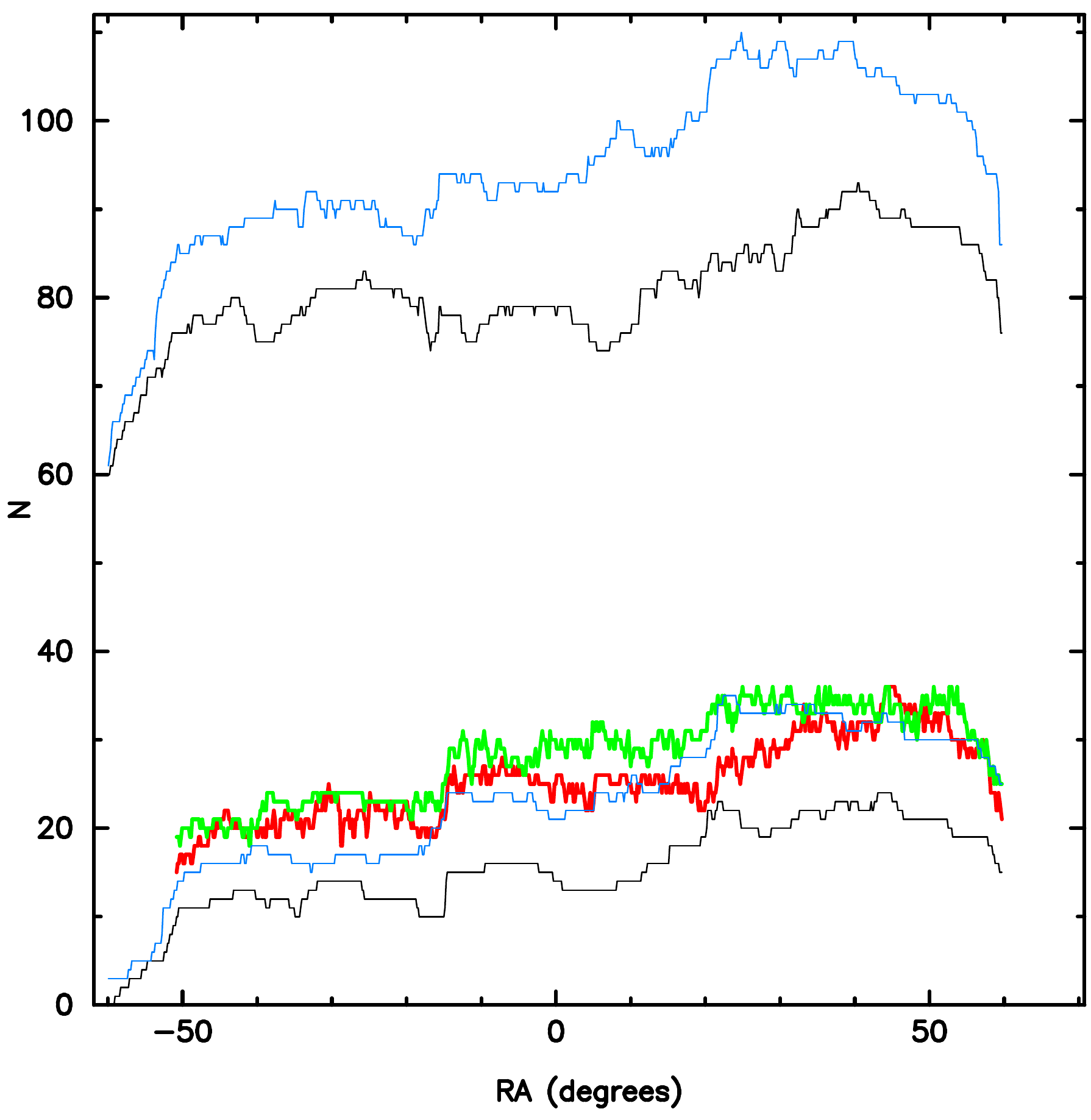}
\caption{
RA distribution of Stripe 82 observations for both runs 106
and 206, corresponding to S and N strips respectively.
The total number of images reaches $\sim 100$ for the S strip 
(blue, top curve) and $\sim 80$ for the N strip (black, top curve).
Nearly $30\%$ of those are calibrated (red and black, bottom 
curves). The number of images selected for the coadd in both N 
(thick green) and S (thick red) strips varies from 15 to 34. 
We therefore expect to see a difference of $\sim 0.4$ mag in depth 
between the shallowest and the deepest regions of the coadd, 
assuming that the signal-to-noise ratio increases as $\sqrt{N}$. 
See text for details on the selection criteria.
}
\label{radecCoverage}
\end{figure}

Figure~\ref{radecCoverage} shows the number of observations as a 
function of RA for runs 106 (S strip) and 206 (N strip) separately. 
The total number of images reaches $\sim 100$ for the S strip 
(blue, top curve) and $\sim 80$ for the N strip (black, top curve).
About $30\%$ of those runs are calibrated (red and black, bottom 
curves) in the sense  
that  the infrared sky camera
indicated a minimum of clouds and a extinction solution was obtained for 
the $20"$ PT telescope data taken at the same night of the run. 
The final number of images used in the coadd, shown as thick green (N strip) 
and red (S strip) lines, varies from 15 to 34. The selection 
criteria to achieve this final sample is described in \S\ref{selectcrit}.
Based on the number of observations used, we 
expect the coadd to be $\sim 2$ mag deeper than 
the single pass data and to show a difference of $\sim 0.4$ mag in depth 
between the shallowest and the deepest regions, 
assuming that the signal-to-noise ratio increases as $\sqrt{N}$.   

\subsection{Field Selection Criteria}\label{selectcrit}

The data selection criteria are listed in Table~\ref{crittable}.
We chose all runs on Stripe 82 with $125 \le$ run $\le 5924$, i.e.,
all data obtained in or before December 1 2005,
demanding that the runs were either on the N or S strip and
rejecting $\sim 10$ runs that were not offset (strip labeled ``O''),
plus one run that is a crossing scan at $\sim 45\arcdeg$ inclination.
We then select the fields in $r$ band, requiring 
seeing better than 2\arcsec, sky brightness 
less than 19.5 mag/arcsec$^2$ and less than 0.2 mag of extinction.
The sky brightness cut corresponds to 2.5 times the median sky of 150 DN,
allowing at most 0.5 mag increase in sky noise.
The majority of the data is uncalibrated so we also require that the 
field has enough stars for our relative calibration method to work.
We cut on the r-band parameters, but rejected all the corresponding
data in $ugiz$. This choice maximizes homogeneity across 
filters, though not optimization for a given filter.

\begin{deluxetable*}{lllc}
\tablecaption{Data Selection Criteria}
\tablehead{Scope & Criterion & Description & Acceptance Rate}
\startdata
\\
run   & $125 \le $ run $\le 5924$         & data taken on or before
                                            12/1/2005 on Stripe 82   & --     \\
\\
field & $r$ $0.265\sqrt{\rm{neff\_psf}} 
        \le 2.0$                          & seeing $<$ 2\arcsec      & 91\%   \\
      & $r$ $\rm{sky\_frames} \le 375$ DN & sky brightness less than 
                                            19.5 mag/arcsec$^2$      & 84\%   \\
      &                                   & 2.5 $\times$ the median  
                                            sky of 150 DN            &        \\
      &                                   & allowing at most 
                                            0.5 mag increase in 
                                            sky noise                &        \\
      & $r$ transparency $> 1/1.2$        & less than 0.2 mag of 
                                            extinction               & 95\%   \\
      & N$_{\rm{calibration}} \ge 1$      & enough stars for 
                                            relative calibration     & 95\%   \\ 
\enddata
\label{crittable}
\end{deluxetable*}

The fraction of fields passing the seeing and sky noise cuts are 91\% and 84\%, 
respectively, and 77.5\% pass both cuts jointly.
However, $100\%$ of the fields in the standard SDSS runs and $95\%$
supernova fields pass 
the transparency cut and the demand to have enough
stars to calculate the photometric scaling.
This indicates that the SDSS had a high
threshold for classifying a night as photometric.
Overall, 1,124,075 frames were included in various fields of the
coadd. The thick red and green lines in Fig.~\ref{radecCoverage}
shows the RA distribution of these selected frames. 

Table~\ref{runtable} summarizes the 123 runs included in the coadd
from both main survey and supernova runs.
69 of these were calibrated (phot=1), in the sense  
that  the infrared sky camera
indicated a minimum of clouds and a extinction solution was obtained for 
the PT telescope data. 
The remaining 54 runs were uncalibrated (phot=0). These were on average 
twice as long as the calibrated ones, as they were taken by the supernova  
survey which was unconcerned with photometricity or sky brightness.

Not all of the images in the available overlapping runs 
were used in a given coadded frame.
In order to prevent sharp discontinuities
in the PSF in the output image at input run edges, we imposed the constraint
that if a field overlaps the output image, 
other good fields from the same run must cover the output image RA completely.

\subsection{Photometric Calibration}
\label{sec:photcal}

The standard SDSS processing calibrates the single pass data using data
obtained by the $20"$ PT telescope. These include a set
of star fields in the Stripe 82 area 
and extinction values measured on the night
the SDSS telescope data are obtained. 
The PT pipeline \citep{Tucker2006} calibration
results in runs calibrated to
rms of $1\%$ in $gri$,   $2\%$ in $u$ and $3\%$ in $z$, \citep{dr1,Ivezic2004}. 
The DR8 data \citep{dr8} includes 
the ubercalibration of \cite{ubercal} which used
Apache Wheel scans and runs obtained for Segue.
The ubercalibration data were not available at the time this work
was performed.

Much of the Stripe 82 data was non-photometric 
so we developed a method to calibrate them. 
In the process
we also re-calibrated those runs taken under photometric conditions.
The runs were calibrated following the prescription of \cite{bramich2008},
which builds a catalog-level coadd of bright stars to
match stars in the frames and compute a zero-point shift.
The resulting photometric calibration of a given run is good to 0.02
mag in up to 1 mag of atmospheric extinction (see also \cite{Ivezic2007}
for a discussion of calibration through clouds).

The PT telescope observed the calibration patches in Stripe 82 many times
while measuring  the extinction for each standard run independently.
Averaging stars calibrated using these independent calibrations
will provide an increase in photometric accuracy. The increase
is unlikely to be $\sqrt{N}$ as there are systematics floors from
residual flat field variations and uncorrected atmospheric transmission
variations.

We used 62 of the photometric runs 
for which normal SDSS PT calibrations were available 
to construct a standard star catalog.
We start with a set
of bright, isolated, unsaturated stars, with $14 < r < 18$, taken from
a set of high quality photometric runs covering both strips of
the whole stripe acquired over an interval of 
less than twelve months (2659, 2662, 2738, 2583, 3325, 3388). 
We then match the individual detections of these stars in each of the
62 runs, using a matching radius of 1 arcsec.  On average,
there are 10 independent measurements of each star among the 62 runs,
and we only include in the reference catalog those with 5 or more
measurements.  
We then compute the mean of the independent calibrated flux 
measurements of each star and adopt that mean flux, defining it 
separately in each band.
We use the fluxes measured in the
SDSS ``aperture 7'', which has a radius of 7.43 arcsec;
this aperture is typically adopted in the SDSS as a reference aperture
appropriate for isolated bright star photometry.

Using this standard star catalog,
we computed the relative zero-point offset of all fields in all runs
used in the coadd, regardless of whether they were photometric or
non-photometric runs initially. The relative zero-point offset was
defined as the median fractional flux difference of the
standard stars in each field in the run.
These field-by-field offsets are the atmospheric transmission $T$, 
which we need for weighting in the coadd 
as well as to place the fields onto the same calibration. 
There is no requirement that $T$ be a smooth function of
RA as for example $T$ will change with time on non-photometric nights. 

The $u$ band images have signal-to-noise ratio significantly poorer 
than the other bands and require special treatment.
All $u$ runs have a provisional calibration applied, but in case of
non-photometric runs these calibrations are
purely the average instrumental zero-point.
We use this approximate calibration to eliminate lower signal-to-noise
stars by rejecting those with $u > 18$. The remaining stars are used
to match against the standard star catalog to find the relative zero-point.

The flux calibrations are relative magnitude 
offsets from a zero-point, $-23.90$. We interpret them
as variations in $T$ with respect to the mean transparency.
We build a table of linear values of these relative flux scale factors $T$.
We $T$ in the coadd image creation process 
to place the images onto the system where
the atmosphere has a uniform transparency, incorporating both
zero-point and extinction from the all-sky-photometry
model of the sky.

\begin{longtable}{rcccccc}
\tablecaption{Runs used in the coadd}
\tablehead{\colhead{run} & \colhead{MJD} & \colhead{date} %
         & \colhead{RA$_{\rm{start}}$}     & \colhead{RA$_{\rm{end}}$} %
         & \colhead{strip} & \colhead{phot} }
  125 & 51081 & 1998/09/25 &  -10.49 &   76.00 & S & 1 \\
 1033 & 51464 & 1999/10/13 &  -49.00 &   -9.40 & N & 1 \\
 1056 & 51467 & 1999/10/16 &  -35.32 &   -0.12 & S & 1 \\
 1752 & 51818 & 2000/10/01 &   21.45 &   79.13 & N & 1 \\
 1755 & 51819 & 2000/10/02 &  -55.68 &   47.42 & S & 1 \\
 1894 & 51875 & 2000/11/27 &   31.68 &   58.91 & S & 1 \\
 2385 & 52075 & 2001/06/15 &  -53.59 &  -37.64 & N & 1 \\
 2570 & 52170 & 2001/09/18 &   17.45 &   59.99 & N & 1 \\
 2578 & 52171 & 2001/09/19 &   29.06 &   61.44 & N & 1 \\
 2579 & 52171 & 2001/09/19 &   36.46 &   60.56 & S & 1 \\
 2583 & 52172 & 2001/09/20 &  -56.42 &  -16.95 & S & 1 \\
 2585 & 52172 & 2001/09/20 &  -32.85 &  -17.43 & S & 1 \\
 2589 & 52173 & 2001/09/21 &   15.79 &   62.58 & N & 1 \\
 2649 & 52196 & 2001/10/14 &  -17.58 &   10.29 & N & 1 \\
 2650 & 52196 & 2001/10/14 &    4.18 &   31.14 & N & 1 \\
 2659 & 52197 & 2001/10/15 &  -58.26 &  -34.58 & N & 1 \\
 2662 & 52197 & 2001/10/15 &  -41.69 &   39.74 & N & 1 \\
 2677 & 52207 & 2001/10/25 &    4.25 &   39.91 & N & 1 \\
 2700 & 52224 & 2001/11/11 &   20.73 &   63.92 & N & 1 \\
 2708 & 52225 & 2001/11/12 &  -15.63 &   25.61 & N & 1 \\
 2709 & 52225 & 2001/11/12 &   20.40 &   63.25 & S & 1 \\
 2728 & 52231 & 2001/11/18 &  -61.21 &   34.24 & N & 1 \\
 2738 & 52234 & 2001/11/21 &   12.86 &   62.18 & N & 1 \\
 2768 & 52253 & 2001/12/10 &  -17.40 &   35.82 & N & 1 \\
 2820 & 52261 & 2001/12/18 &   20.66 &   61.32 & N & 1 \\
 2855 & 52282 & 2002/01/08 &   19.76 &   30.66 & N & 1 \\
 2861 & 52283 & 2002/01/09 &   32.84 &   66.17 & N & 1 \\
 2873 & 52287 & 2002/01/13 &   14.21 &   62.59 & N & 1 \\
 2886 & 52288 & 2002/01/14 &   14.21 &   62.52 & S & 1 \\
 3325 & 52522 & 2002/09/05 &  -15.54 &   61.16 & S & 1 \\
 3355 & 52551 & 2002/10/04 &   19.37 &   61.00 & S & 1 \\
 3360 & 52552 & 2002/10/05 &  -53.45 &   25.70 & S & 1 \\
 3362 & 52552 & 2002/10/05 &   20.47 &   57.45 & N & 1 \\
 3384 & 52557 & 2002/10/10 &  -54.65 &   66.50 & N & 1 \\
 3388 & 52558 & 2002/10/11 &  -47.22 &   62.66 & S & 1 \\
 3427 & 52576 & 2002/10/29 &  -51.41 &  -24.96 & S & 1 \\
 3430 & 52576 & 2002/10/29 &   20.76 &   40.16 & S & 1 \\
 3434 & 52577 & 2002/10/30 &  -52.22 &   36.25 & S & 1 \\
 3437 & 52578 & 2002/10/31 &  -50.78 &   24.99 & N & 1 \\
 3438 & 52578 & 2002/10/31 &   30.63 &   62.61 & S & 1 \\
 3460 & 52585 & 2002/11/07 &   19.41 &   61.44 & S & 1 \\
 3461 & 52585 & 2002/11/07 &   42.77 &   61.24 & N & 1 \\
 3465 & 52586 & 2002/11/08 &  -34.17 &   21.68 & S & 1 \\
 4128 & 52908 & 2003/09/26 &   -7.82 &   61.45 & N & 1 \\
 4136 & 52909 & 2003/09/27 &   27.90 &   60.79 & S & 1 \\
 4145 & 52910 & 2003/09/28 &  -16.01 &   61.85 & S & 1 \\
 4153 & 52911 & 2003/09/29 &  -16.34 &   11.87 & N & 1 \\
 4157 & 52912 & 2003/09/30 &   19.23 &   61.20 & N & 1 \\
 4184 & 52929 & 2003/10/17 &  -52.92 &  -10.08 & N & 1 \\
 4187 & 52930 & 2003/10/18 &  -51.72 &  -34.03 & S & 1 \\
 4188 & 52930 & 2003/10/18 &  -15.85 &    8.25 & N & 1 \\
 4192 & 52931 & 2003/10/19 &  -52.66 &   23.32 & S & 1 \\
 4198 & 52934 & 2003/10/22 &  -53.56 &   61.16 & N & 1 \\
 4203 & 52935 & 2003/10/23 &  -60.00 &   61.51 & S & 1 \\
 4207 & 52936 & 2003/10/24 &  -55.00 &   61.38 & N & 1 \\
 4247 & 52959 & 2003/11/16 &  -15.67 &   28.74 & S & 1 \\
 4253 & 52962 & 2003/11/19 &  -15.59 &   13.24 & N & 1 \\
 4263 & 52963 & 2003/11/20 &  -16.70 &   53.98 & S & 1 \\
 4288 & 52971 & 2003/11/28 &   19.27 &   46.65 & S & 1 \\
 4797 & 53243 & 2004/08/26 &  -53.53 &  -24.28 & N & 1 \\
 4868 & 53286 & 2004/10/08 &  -30.53 &   62.90 & N & 1 \\
 4874 & 53288 & 2004/10/10 &  -62.40 &   88.08 & N & 1 \\
 4895 & 53294 & 2004/10/16 &   -4.80 &   70.77 & N & 1 \\
 4905 & 53298 & 2004/10/20 &    0.38 &   72.29 & N & 1 \\
 4917 & 53302 & 2004/10/24 &  -65.59 &   52.81 & N & 0 \\
 4930 & 53313 & 2004/11/04 &  -58.01 &    1.81 & S & 1 \\
 4933 & 53314 & 2004/11/05 &  -53.70 &   63.36 & N & 1 \\
 4948 & 53319 & 2004/11/10 &    7.31 &   62.39 & N & 1 \\
 5042 & 53351 & 2004/12/12 &   18.42 &   61.43 & S & 1 \\
 5052 & 53352 & 2004/12/13 &  -15.67 &   25.72 & S & 1 \\
 5566 & 53616 & 2005/09/03 &  -33.61 &   60.28 & N & 0 \\
 5582 & 53622 & 2005/09/09 &  -55.62 &   58.95 & S & 0 \\
 5590 & 53623 & 2005/09/10 &  -60.69 &   12.78 & N & 0 \\
 5597 & 53625 & 2005/09/12 &  -64.68 &  -17.02 & S & 0 \\
 5603 & 53626 & 2005/09/13 &  -66.47 &   63.02 & N & 0 \\
 5607 & 53627 & 2005/09/14 &  -63.90 &   62.25 & S & 0 \\
 5610 & 53628 & 2005/09/15 &  -66.73 &   65.38 & N & 0 \\
 5619 & 53634 & 2005/09/21 &  -64.44 &   63.20 & S & 0 \\
 5622 & 53635 & 2005/09/22 &  -64.48 &   63.38 & N & 0 \\
 5628 & 53636 & 2005/09/23 &  -64.61 &   21.61 & S & 0 \\
 5633 & 53637 & 2005/09/24 &  -61.59 &   59.71 & N & 0 \\
 5637 & 53638 & 2005/09/25 &  -21.61 &   63.42 & S & 0 \\
 5642 & 53639 & 2005/09/26 &  -10.71 &   62.83 & N & 0 \\
 5646 & 53640 & 2005/09/27 &  -65.64 &   70.65 & S & 0 \\
 5658 & 53641 & 2005/09/28 &   15.19 &   56.03 & N & 0 \\
 5666 & 53643 & 2005/09/30 &   40.36 &   63.58 & S & 0 \\
 5675 & 53645 & 2005/10/02 &  -60.60 &  -40.41 & S & 0 \\
 5681 & 53646 & 2005/10/03 &   22.34 &   54.21 & S & 0 \\
 5709 & 53654 & 2005/10/11 &  -67.46 &   24.71 & N & 0 \\
 5713 & 53655 & 2005/10/12 &  -68.40 &   42.80 & S & 0 \\
 5731 & 53657 & 2005/10/14 &   20.28 &   62.34 & N & 0 \\
 5732 & 53657 & 2005/10/14 &   46.22 &   62.33 & S & 0 \\
 5754 & 53664 & 2005/10/21 &  -57.68 &   59.33 & S & 0 \\
 5759 & 53665 & 2005/10/22 &  -59.55 &   59.24 & N & 0 \\
 5763 & 53666 & 2005/10/23 &  -59.22 &    5.29 & S & 0 \\
 5765 & 53666 & 2005/10/23 &    1.41 &   56.19 & N & 0 \\
 5770 & 53668 & 2005/10/25 &  -56.57 &   59.20 & N & 0 \\
 5771 & 53668 & 2005/10/25 &   32.20 &   62.16 & S & 0 \\
 5776 & 53669 & 2005/10/26 &  -59.98 &   59.26 & S & 0 \\
 5777 & 53669 & 2005/10/26 &   23.52 &   59.29 & N & 0 \\
 5781 & 53670 & 2005/10/27 &  -56.20 &   59.17 & N & 0 \\
 5782 & 53670 & 2005/10/27 &   31.77 &   62.19 & S & 0 \\
 5786 & 53671 & 2005/10/28 &  -36.66 &   63.30 & S & 0 \\
 5792 & 53673 & 2005/10/30 &  -62.44 &   59.30 & N & 0 \\
 5797 & 53674 & 2005/10/31 &  -59.00 &   59.39 & S & 0 \\
 5800 & 53675 & 2005/11/01 &  -59.49 &   59.98 & N & 0 \\
 5807 & 53676 & 2005/11/02 &  -48.55 &   59.20 & S & 0 \\
 5813 & 53677 & 2005/11/03 &  -65.13 &   45.97 & N & 0 \\
 5820 & 53679 & 2005/11/05 &  -45.43 &   62.22 & S & 0 \\
 5823 & 53680 & 2005/11/06 &  -60.09 &   62.32 & N & 0 \\
 5836 & 53681 & 2005/11/07 &  -60.56 &   62.35 & S & 0 \\
 5842 & 53683 & 2005/11/09 &  -59.31 &   62.22 & N & 0 \\
 5847 & 53684 & 2005/11/10 &  -63.50 &   63.90 & S & 0 \\
 5866 & 53686 & 2005/11/12 &   17.32 &   63.41 & N & 0 \\
 5878 & 53693 & 2005/11/19 &  -62.67 &   64.29 & N & 0 \\
 5882 & 53694 & 2005/11/20 &  -63.69 &   63.13 & S & 0 \\
 5889 & 53696 & 2005/11/22 &   35.34 &   63.13 & S & 0 \\
 5895 & 53697 & 2005/11/23 &  -63.25 &   62.62 & S & 0 \\
 5898 & 53698 & 2005/11/24 &  -65.81 &   62.31 & N & 0 \\
 5902 & 53699 & 2005/11/25 &  -62.81 &   62.75 & N & 0 \\
 5905 & 53700 & 2005/11/26 &  -68.06 &   62.74 & S & 0 \\
 5918 & 53704 & 2005/11/30 &  -62.39 &   62.28 & N & 0 \\
 5924 & 53705 & 2005/12/01 &  -63.20 &   62.51 & S & 0 
\label{runtable}
\end{longtable}


\section{Coadd Image Creation}\label{imagecreation}

We aim at coadding the Stripe 82 data and running it through \photo.
To make the coadd we need the data images and weight maps.
We build a weight map by multiplying an inverse variance map and 
a geometry mask.
\photo\ requires a map of the saturated pixels. 
In this section  detail the process of
creation of each of these image components and describe how we use them in 
the coaddition.

\subsection{Sky Subtraction}
The sky brightness varies both spatially and temporally, night to night
and due to clouds. We removed the sky before mapping. 
As the data we used for the coadd came from DR7 and earlier,
the improved sky subtraction of the current DR8 \photo\ was not implemented
(\cite{dr8}; see also \cite{blanton2011} for continued work
on this subtle problem).
The DR7 \photo\  algorithm, which was current at the time, produced a sky 
image by
calculating  the median of the $256\times256$ pixel boxes
in the data image on a grid of 128 pixels in each dimension.
The sky was then determined using bilinear interpolation.
This algorithm over-subtracts the extended parts of galaxies 
on the scale of the $128\times128$ pixel
grid used in the sky calculation, leaving artifacts due to astrophysical
objects in the data.
\photo\ sky subtraction
engine was therefore deemed not suitable for our purposes and we 
developed our own method.

Subtracting a global sky for each frame would work poorly as the sky changed
with time and thus with row number in the SDSS data.
Instead we adopted a sky value that was allowed to vary
linearly with row number and thus time. 
For a given frame and the two frames on either side of it in the run
(and after removing the SDSS soft bias of 1000 DN),
the median of the 2048 pixels along each row was calculated,
resulting in a $3\times1489$ pixel sky vector.  
This vector is a time series estimate. 
In order to deal with bright stars,
the rms of the vector was calculated using a $3\sigma$ 
5 iteration sigma clipping, and the rms
was used to reject pixels more than $2\sigma$ from the mean.
A linear least squares fit to the
remaining vector was used to model the sky, which was subtracted from the image
row by row.

\subsection{Astrometry}\label{sec:astrom}

The coadd relies on the existing astrometry produced using the \astrom\
pipeline \citep{astrom}.
All of the images used in the coadd had astrometric calibrations.
\cite{astrom} were able to achieve positions accurate to
$\sim45$ mas rms per coordinate by calibrating
to the US Naval Observatory CCD Astrograph Catalogue (UCAC; \cite{ucac}).  
The accuracy is limited primarily by the accuracy of the UCAC positions 
($\sim70$ mas rms at the UCAC survey limit of $R \approx 16$)
and the density of UCAC sources. This accuracy can be represented
in the affine transformations that are standard in the WCS convention.
\cite{astrom} also noted that there are systematic optical 
distortions due to the camera 
present in the data. We will use the \astrom\ measurements to 
remove these distortions. 

The process of forward mapping requires a transformation from RA,Dec 
($\alpha$,$\delta$) to pixel
location in the input image.
The SDSS runs were taken along great circles.
Thus \astrom\ worked in a coordinate system 
in which each run's great circle is the equator of the coordinate system.  
In this great circle coordinate system, 
the latitude of an observed star never exceeds about $1.3 \arcdeg$;
thus the small angle approximation may be used and lines of
constant longitude are, to an excellent  approximation, perpendicular to lines of
constant latitude.  Longitude and latitude in great circle coordinates are
referred to as $\mu$ and $\nu$, respectively.  $\nu$ is equal to 0 along
the great circle, $\mu$ increases in the scan direction, and the origin of
$\mu$ is chosen so that $\mu = \alpha_{2000}$ at the ascending node (where
the great circle crosses the J2000 celestial equator).
The conversion from great circle coordinates to J2000 celestial coordinates
is then
\begin{eqnarray}\label{greatcirc}
\tan{(\alpha-\mu_0}) &=& \frac{\sin(\mu - \mu_0) 
\cos \nu \cos i - \sin \nu \sin i }{\cos(\mu - \mu_0) \cos \nu} \\
\sin{(\delta)} &=&  \sin (\mu - \mu_0) \cos \nu \sin i + \sin \nu \cos i 
\end{eqnarray}
where $i$ and $\mu_0$ are the inclination and J2000 right ascension of the
great circle ascending node, respectively.
$\mu_0 = 95\arcdeg$ for all survey stripes, and for Stripe 82 $i \approx 0$.

Given  the great circle coordinates ($\mu$, $\nu$) we can transform to 
distortion corrected frame coordinates ($x'$,$y'$) using the affine 
transformation
\begin{eqnarray}\label{affine}
\mu_{CMP} & = & a + b x' + c y' \\
\nu_{CMP} & = & d + e x' + f y'
\end{eqnarray}
The transformation from ($x'$, $y'$) to ($x$, $y$) accounts for optical
distortions which, in drift-scan mode, are a function of column only: 
\begin{eqnarray}
& & x' = x + g_0 + g_1 y + g_2 y^2 + g_3 y^3 \\
& & y' = y + h_0 + h_1 y + h_2 y^2 + h_3 y^3   \label{distort}
\end{eqnarray}
Equations~(\ref{greatcirc}-\ref{distort}) provide the pixel coordinates 
on the input image that corresponds to a given (RA,Dec) position.

\cite{bramich2008} performed a recalibration of the
Stripe 82 astrometry using a run from the mid-time
of the Stripe 82 observations as a reference. This
removed an erroneous but measurable galaxy 
mean proper motion of $\sim10$ mas/yr
in both RA and Dec due the proper motion of reference
stars (see \cite{bramich2008}, Fig.~4). 
The optics distortions removed in this section have a maximum peak to
peak shift of 80 mas (see \cite{astrom}, Fig.~2)
and  are larger than the
astrometry shifts induced by stellar motion over the 5 year range 
of the data used in the coadd.
Therefore, although ideally we should have removed the reference star proper 
motion drift, in practice these have little effect.

\subsection{Astrometric Mapping of Data Images}

We geometrically map the input images onto the output image.
We defined output frames aligned along the J2000
equator with rows aligned perpendicular to RA,
in a standard SDSS image format, from
$-50\arcdeg \le \alpha \le 60\arcdeg$ and
$-1.25\arcdeg \le \delta \le 1.25\arcdeg$. 

Since the output image is simply a locally flat tangent projection
of the sky,  the mapping must remove optical distortions and 
provide a surface brightness estimate at the aligned pixel location.
To perform the mapping we used a version of Swarp \citep{swarp} 
modified to perform the astrometric conversions described 
in~\S\ref{sec:astrom}.

Each pixel in the mapped image is estimated from
the input image pixels using a Lanczos interpolation kernel, which is
a truncated sinc interpolation. 
Given bandwidth limited signal of infinite
extent, sinc function interpolations reproduce exactly
the data after resampling.
Our data is not undersampled (it has 0.4\arcsec\ pixels and seeing of
$\sim 1.3\arcsec$), but it is not of infinite bandwidth either and 
this motivates the use of a truncated kernel.

We used a two dimensional Lanczos-3 kernel retaining 3 maxima on each
side of the center in each dimension.
The one dimensional Lanczos-3 kernel is
$ L(x) = sinc(x)sinc(x/3) $, for $-3 < x  < 3$. 
Then the 2 dimensional interpolation formula is
\begin{equation}\label{Lanczos}
\hat{I}(r,c)  = \sum_{i,j} I(i,j) L(r-i) L(c-j)
\end{equation}
where $r,c$ are the output image pixel coordinate and $i,j$ are the
input image pixel coordinate.
The Lanczos-3 window is well-behaved in terms 
of reduction of aliasing, minimal ringing, and lack of smoothing, but
a Lanczos-3 interpolation does use $(2x3+1)^2 =49$ pixels 
to estimate the value of one output pixel. This is too large
to use for bad pixels (e.g, saturated pixels), so we use
nearest neighbor interpolation and reduce the weight of bad pixels
during Lanczos-3 interpolation by using a mask 
(see section~\S\ref{sec:astroMaskMap}).

\subsection{Inverse Variance Map}

To keep track of the variance of  data images, 
pixel by pixel inverse variance images are 
a natural choice, but they produce biases in the resulting mean.
At low signal to noise the upward fluctuations in signal
are given more weight than downward fluctuations as a result
of the one-sided nature of the Poisson distribution.
This bias is deterministic and one could
correct for it, but, for example, for $u$-band data with its $120e^-$
of sky noise, pixels at $1\sigma$ above sky would be biased
by $0.5\%$ and this would be a fair fraction of our photometric
error budget. Another problem is that per-pixel inverse
variance weighting systematically changes the shape
of the PSF as a function of the magnitude of the object. This 
would cause serious complications to our PSF-based photometry 
using \photo. For these reasons we chose a different method.

We computed the variance of the sky
as measured on the frame from the width of the sky histogram.
As we used a linear gradient sky subtraction, each
image is assigned a variance image that is
the variance of the subtracted linear sky gradient. 
This of course assumes  Poisson statistics while the data are in ADU.
In calculating this variance we did not include
the effective gain, $g_{\rm eff}$, so  
gain variations are not accounted for. These variations are $<30\%$ though.

\subsection{Geometry Mask}
\label{sec:geoMask}

The geometry mask, keeps track of which pixels in the input image 
actually contribute to the coadd.
In this mask definition we account for image defects found by \photo, 
in particular, cosmic rays, saturated pixels and bad columns.

\photo\ produces a 16 bit mask image, the {\tt fpM} file.
For the geometry mask we are interested in the INTERP bit.
The INTERP flag is set for any pixel
for which \photo\ used interpolation to fill its value.
This happens for cosmic rays, saturated pixels, and bad columns.
These pixels are poor but not useless estimates of the true value 
of the pixel. We set the geometric weight of such pixels 
to a small value, 0.0000001.
This ensures that if there is no input image which contributes a good
estimate for a given output frame pixel, such as the center of a saturated
star, the interpolated value is used.

The geometry of the SDSS images are also encoded into the geometry mask.
In SDSS images, the first 124 rows of each frame are duplicates of the last 124
rows of the preceding frame. This replication of pixels was done 
so that objects could be well measured despite being on edge of a frame. 
To account for this, the first 124 rows of the mask are set to 0. 
In the SDSS images there are also scanline to
scanline overlaps between North and South strips. This is
actual exposure time, again designed to make objects on 
the edges measurable. For our purposes we kept the North and South
strip data separate as the extra exposure was on too thin a strip to
be useful.

\subsection{Satur Mask}
\label{sec:saturMask}

\photo's bit mask also contains information about saturated pixels, 
encoded in the SATUR bit. The SATUR flag is set for any pixel that  \photo\ 
determines to be saturated.
We set these pixels to 1 in an image otherwise filled with zeros.
This is the SATUR map, which we need for running \photo\ on the 
coadded images. 

\subsection{Weight Map} \label{sec:weightMap}

Although closely related, 
the weigh map and the inverse variance image are not the same.
The weight map is the inverse
variance image multiplied by the geometry mask.
We set all masked non-zero pixels to the INTERP flag value (0.0000001). This 
allows us to keep track of the pixels altered by the masking.

We  use the weigh map as input for the coaddition process,
in which a weighted clipped mean of the data images, all mapped onto the 
same output image, will be performed.
In addition, we  coadd the inverse variance images and the satur maps as well,
and although for those we use straight sums, the weigh map plays a role in the 
sense that only the pixels that pass the clipping for the data coadd are included
(for details, see section~\S\ref{sec:wcm}).


\subsection{Astrometric Mapping of Map \& Mask Images}
\label{sec:astroMaskMap}

The inverse variance map, satur mask and weight map are all 
mapped onto the same output image as the data.
For the inverse variance and weight images we apply the same Lanzcos-3 
interpolation used for the data in order to replicate the noise correlation.
The satur masks, propagating saturated pixels, 
were mapped using a nearest neighbor interpolation.
This is more suitable than Lanzcos-3 
because masked pixels an area of influence limited to their first neighbors.

In the process of performing the wide Lanzcos-3 interpolation of
the data, the geometry mask is used to prevent masked pixels from
contributing to the interpolation with more than minimal weight.
In addition, once the images are mapped onto the output image, any non-overlap is
set to zero weight using the 
 geometry mask.

After this mapping procedure, each stack of images is ready for the 
coaddition process.
No scaling has been done at this stage, so
the stack  for a given output image can be 
further filtered and/or  weighted as needed.

\subsection{Coaddition}

With the four stacks of images (data, inverse variance, weight and satur)
all aligned and cropped to match the output image, 
we have all the inputs needed to produce the coadded images  
and run \photo\ on them. We chose to use all the data in our image set.
This allows us to maximize the depth of the coadd data. 

Alternatively,
one could devise a selection criteria to produce a coadd dataset tailored 
for a specific purpose. 
For example, one could select only images with the best
seeing and limit them in number to form a uniform depth, aiming at 
weak lensing studies. 

\subsubsection{Weights}
\label{sec-wts}

The data in our image set is of variable quality and we wish
to optimize our coadd, so we designed a weighting scheme.
The signal to noise ratio of the measurement of flux from a star is:
\begin{equation}
S/N \propto {{N_{\rm photons}}\over{\sqrt{A_{\rm psf}} \sigma_{\rm sky}}} 
\end{equation}
where $N_{\rm photons}$ is the number of photons detected from the source,
$A_{\rm psf}$ is the area in pixels the source subtends, and $\sigma_{sky}$
is the sky noise per pixel.
As $N$ is proportional to transparency $T$ and $A \propto {\rm FWHM}^2$,
we use the following as weights:
\begin{equation}\label{eq-weights}
w_i = {{T_i}\over{{\rm FWHM_i}^2 \sigma_i^2(p_i)}}
\label{eq-weights}
\end{equation} 
This gives highest weight to good
seeing data taken when the sky is clear and dark.
We choose these weights because although the usual inverse variance weighting
produces the minimum variance image, the  signal to noise
for a star depends on square of seeing.
With this weight, the PSF of the coadd is $0.3\arcsec$ less than
the median of the input images in $ugri$
and $0.2\arcsec$ less in $z$.

We take the seeing for each field and filter from the {\tt tsField} file,
which is simply an average over the frame, and use it regardless of how much the 
frame contributes to a given output coadd frame. The transparency $T$, in turn,
is already available as a product of the photometric calibration discussed 
in section~\S\ref{sec:photcal}.
We have therefore cataloged for each image the transpareancy, the seeing, 
and via the inverse variance image, the sky variance. 
This allows us to proceed to the coaddition.

\subsubsection{Weighted Clipped Mean}\label{sec:wcm}
We coadd by building the stack of mapped images on a given output frame
and performing a weighted clipped mean. 
The weight images described in \S\ref{sec:weightMap} are multiplied
by $T/{\rm FWHM}^2$. Flux calibration of our data images by $1/T$
means that our weight images get another factor of $T^2$ due to error
propagation.

Then for each output pixel we collect the corresponding mapped pixels, 
and reject outliers,
using a iterative $5\sigma$ rejection, where for ``$\sigma$'' we use the
25-75\% interquartile range 
The final average uses the weights of Eq.~\ref{eq-weights}.

The data images are coadded as described above.
The inverse variance image and the satur mask coadds were computed using
straight sums (not averages), though using only
those pixels corresponding to those that make it past the clipping process
into the coadded data image.

\subsection{Output Images}
\begin{figure*}
\includegraphics[width=1.0\linewidth]{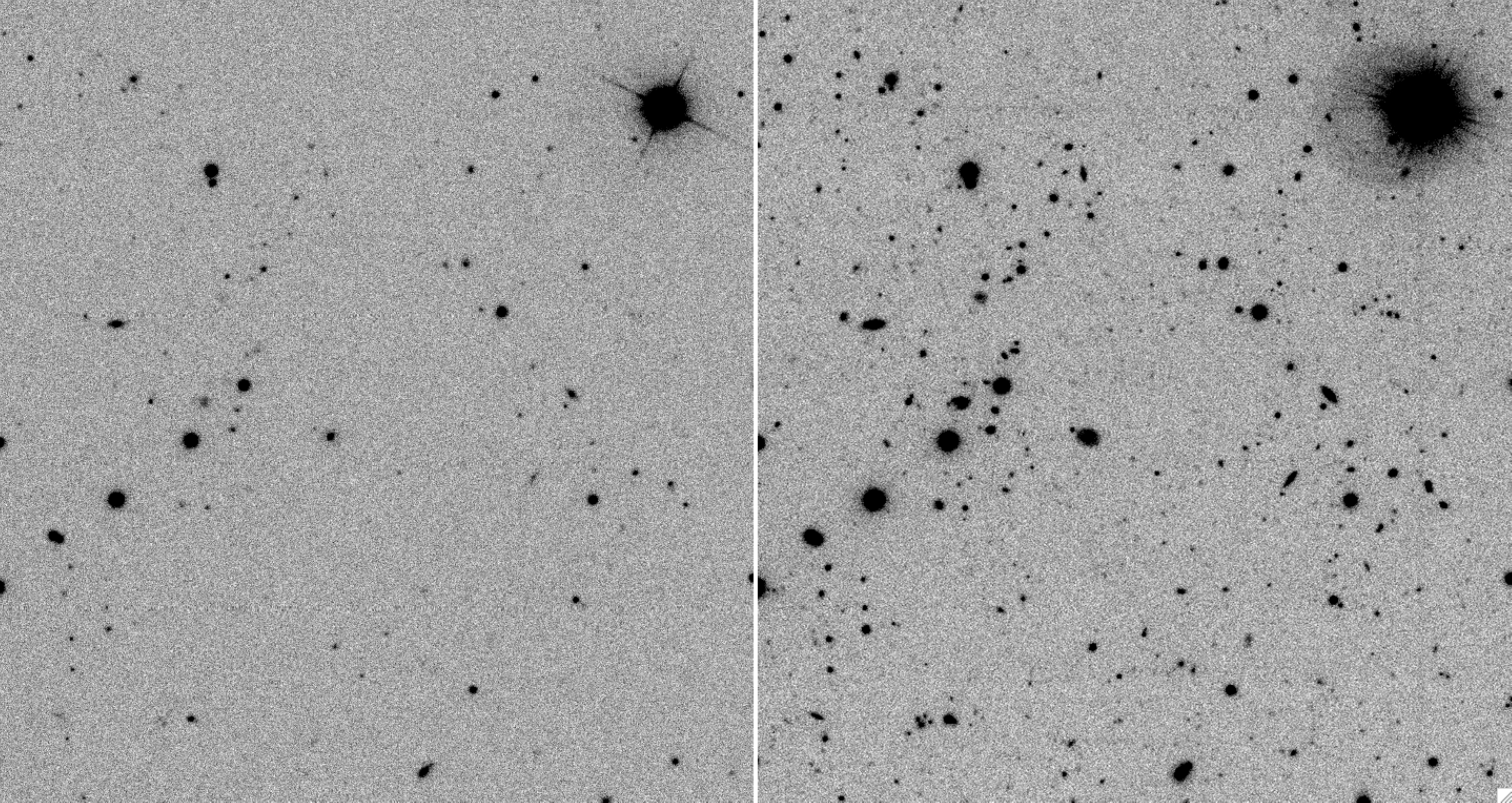}
\caption{
Comparison between single pass (left) and coadd (right) images in $r$-band 
for run 206, camcol 3, field 505, RA=15, Dec=0. 
Images are shown with the same 
scale, contrast and stretch.
The single pass counterpart (run 5800, camcol 3, field 505) 
is one out of 28 images used in the  coaddition
of this particular image.  
This example illustrates the fact that a large 
number of objects bellow the detection 
threshold of each image can be well detected and measured in the coadd.
}
\label{singleCoadd}
\end{figure*}

Figure~\ref{singleCoadd} shows a side by side comparison between  coadd 
single pass data in $r$-band. The single pass counterpart is one out of 28 
images used in the coaddition
which means that it passed all the cuts discussed in the text.  
This example features run 206, camcol 3, field 505 in $r$-band.
It illustrates the fact that a number of objects bellow the detection 
threshold of each image can be well detected and measured in the coadd.
The seeing, however, is larger for the coadd image in this example, 
indicating that this particular single pass image has seeing better then
the median seeing on the stack. 

\section{Catalog Creation}\label{catalogcreation}
The coadd production yields 800 fields along six columns in each of two stripes
and each field includes data in five bands. The resulting 48,000 images 
are processed through a modified version of the SDSS image processing
pipeline \photo, to yield object catalogs. 

\subsection{PSF Measurement}\label{sec:psf}
Our principal challenge is
to measure the PSF of the coadd.
The PSF of SDSS images varied in time, 
corresponding to image rows, due
to atmospheric fluctuations. It also  varied in space, 
corresponding to columns, due to camera optics. 
In the standard single run reduction, \photo\
fits the spatial variation of the PSF by 
computing a PSF basis set using a Karhunen-Lo\'{e}ve (KL) expansion. 
The stars in the frame and the two flanking on either side (five in total)
are used to determine the KL basis functions $B_R(u,v)$:
\begin{equation}\label{eq-kl-basis}
P_{i}(u,v) = \sum_{r=1}^{r=n} a^r_{i} B_R(u,v)
\end{equation}
where $P_{i}$ is the PSF of the $i^{th}$ star,  $u,v$ are the pixel coordinates
relative to the basis function origin, and $n$ sets the number of 
terms to use in the expansion (we use $n=3$).
The stars in the image and the flanking half an image on either side (two in total)
are used to determine the KL coefficents $a^r_{i}$:
\begin{equation}\label{eq-kl-coeffs}
 a^r_{i} \approx \sum_{l=m=0}^{l+m\le N} b^r_{lm}x^l_{i}y^m_{i}
\end{equation}
where $x,y$ are the coordinates of the center of the $i^{th}$ star, $N$
is the highest power of $x$ or $y$ included in the expansion (we use $N=2$,
a quadratic spatial variation of the PSF), and the $b^r_{lm}$ are found
by minimising
\begin{equation}
\sum_i (P_{i}(u,v) - \sum_{r=1}^{r=n} a^r_{i} B_r(u,v))^2 
\end{equation}

In the coadd we  took advantage of the fact that the KL basis set
and coefficients had already been determined
each input image as each of the input runs had already been
separately processed through \photo.
Using the KL basis set for each input run corresponding to the given
output frame and the two flanking frames,
we compute a weighted sum of the model PSFs (using the
weights of equation~\ref{eq-weights}) on a 2-dimensional grid with a spacing
of $\sim 1.5\arcmin$.  We then fit the KL basis $B_R(u,v)$'s to these over the
coadd frame and the two flaking frames.
The computation of the coefficents of the PSF expansion $a^r_{i}$ are
done on the coadd frame and the two flanking half-frames
as in the single pass runs, with the exception that we set
the maximum number of
coefficients, $n$ in Eq.~\ref{eq-kl-basis}, to $n = 4$ 
and highest power of $x,y$, $N$ 
in Eq.~\ref{eq-kl-coeffs}, to $N = 3$.


\subsection{Effective Gain \& Sky}
\photo\ had to be modified to read in a file containing the weights and
effective gains of the images.
The effective gain of a coadd image $G$ should be such that it gives
bright objects Poisson statistics for the variance associated with
a source when scaling the averaged pixel counts back 
to electrons. Therefore, 
\begin{equation}\label{eq:PoverG}
\sigma^2(P) = P/G
\end{equation}
 where $P$ is the weighted sum, over all exposures $i$,  
of the object counts $p_i$, 
\begin{equation}
P = \sum w_ip_i = \sum w_i T_i o_i
\end{equation}
and $p_i \equiv T_i o_i$ is the object's flux per exposure $o_i$ times 
the flux scale factor $T_i$. The corresponding variance is:
\begin{eqnarray}
\nonumber \sigma^2(P) & = &  \sum w^2_i\sigma^2(p_i)       \\ 
\nonumber             & = &  \sum w^2_i T^2_i\sigma^2(o_i) \\
\nonumber             & = &  \sum w_i^2T^2_io_i/g_i        \\
                      & = &  \sum p_iw_i^2T_i/g_i
\end{eqnarray}
Assuming that $p_i$ does not vary much from image to image, we obtain:
\begin{equation}\label{eq:Approx}
\sigma^2(P) \approx \langle p_i \rangle \sum w^2_i T_i/g_i = P \sum w^2_i T_i/g_i
\end{equation}
Eqs.~\ref{eq:PoverG} and \ref{eq:Approx} we conclude that the effective gain is:
\begin{equation}
G = \left[ \sum w_i^2T_i/g_i\right]^{-1} 
\end{equation}
In the SDSS the gains $g_i$ of all frames going into a single coadd frame are 
the same (it is always the same CCD),
so they may be replaced by a single $g$ outside the sum:
\begin{equation}\label{eq:EffGain}
G = \frac{g}{\sum w_i^2T_i}
\end{equation}

To process the images with \photo\ we also need to compute the effective sky 
level $S'$. Assuming that dark and read 
noise are accounted for in the Poisson noise of 
the effective sky, we can easily obtain  $S'$ using 
the effective gain $G$ calculated in Eq.~\ref{eq:EffGain} and the relation
$S'=G\sigma^2(S')$, which is analogous to Eq.~\ref{eq:PoverG}.  
We  used the clipped variance of the coadd frame to compute $\sigma^2(S')$.
Alternatively, one could take the sky variances and weights of the input
frames and compute it using $\sigma^2(S')=\sum w_i^2T_i^2\sigma^2(s_i)$.

Using this method  we obtained, for each field and for all 5 filters, 
the effective sky ($S'$), sky noise ($\sigma(S')$) and gain 
($G$, Eq.~\ref{eq:EffGain}). With these quantities in addition to 
the weights ($w_i$, Eq.~\ref{eq-weights}) we processed all of the coadd
fields through \photo.

\subsection{Applying the Calibration}\label{sec:target}

The SDSS code {\tt Target} is used to apply the calibration to the raw outputs
of \photo. Following \cite{lupton1999}, we convert from fluxes $f$ to mags $m$ using
\begin{equation}
\label{luptitude}
m(f/f_0) = -\frac{2.5}{\ln 10} \left[{\rm asinh}\left(\frac{f/f_0}{2\,b}\right) 
       +\ln(b)\right] 
\end{equation}
where $b$ is an arbitrary dimensionless 
softening parameter below which the magnitude scale goes
from linear to logarithmic and $f_0$ is a reference flux that sets the 
zero-point, $zp \equiv m(0)$, of the magnitude scale.
The values of $b$ that we used for the
 coaddition is given in  Table~\ref{table:asinh}, along with the asinh
 magnitudes associated  with a zero flux object and
the magnitudes  corresponding to $f=10f_0 b$. Above this scale, the asinh
magnitude and the traditional logarithmic magnitude differ by less
than 1\% in flux. These values can be compared to their  
 equivalent numbers for the main survey, given in Table 21 of \cite{edr}.
The coadd images were all placed onto a 
uniform flux scale such that 1 DN corresponds to a flux of 
1 picomaggie, corresponding to a logrithmic (and not asinh) mag of 30.

\begin{deluxetable}{cccc}
\tablecaption{Asinh Magnitude Softening Parameters for the Coadd\label{table:asinh}}
\tablecolumns{4}
\tablehead{\colhead{filter} & \colhead{$b$} & \colhead{$zp$} & \colhead{$m(10b)$}}
\startdata
$u$ & $1.0\phn \times 10^{-11}$ & 27.50 & 24.99\\
$g$ & $0.43    \times 10^{-11}$ & 28.42 & 25.91\\
$r$ & $0.81    \times 10^{-11}$ & 27.72 & 25.22\\
$i$ & $1.4\phn \times 10^{-11}$ & 27.13 & 24.62\\
$z$ & $3.7\phn \times 10^{-11}$ & 26.08 & \phn23.57
\enddata
\tablecomments{Values reported by \cite{dr7}. Column $zp$ is 
the zero-point magnitude, $zp \equiv m(0)$. The final column gives 
magnitude associated with an object for which 
$f/f_0 = 10b$.}
\end{deluxetable}

\subsection{Star/Galaxy Separation}
\label{stargalx}
The SDSS standard star/galaxy separation simply classifies
as stars all objects in the region \citep{dr2}
$|r_{\rm{psf}} - r_{\rm{model}}| \le 0.145 $ 
where $r_{\rm{model}}$ is the model magnitude (the best fit galaxy deVaucouleurs or 
exponential profile convolved with the PSF) and $r_{\rm{psf}}$ is the 
PSF magnitude at the position of the object. 
This simple estimator performs well for single pass data:
$95\%$ correct at magnitude 21 in $r$.

In the case of the coadd data, star count plots showed dramatic increases
in the numbers of stars at magnitudes where galactic models do not.
This suggested that more numerous galaxies were being misclassified as stars there.
Thus we instead used the more stringent criterion
\begin{equation}
|r_{\rm{psf}} - r_{\rm{model}}| \le 0.03 
\end{equation}
to select stars. As \photo\ measures
every parameter for every object regardless of its
determination of object type this has no effect on the
other measurements.


\section{Data Products Verification}\label{results}

\subsection{Photometric Calibration}

We use the standard star catalog of 
\cite{Ivezic2007} to verify our photometric calibration.
To build that catalog \cite{Ivezic2007} took the median of individual 
measurements of bright stars from 58 Stripe 82 runs and then applied several  
corrections to their catalog: $1)$ a color-term like correction
for the bandpass of each camera column; $2)$ a purely RA flat field correction
in $r$ derived by comparing PT data with SDSS data; and
$3)$ a purely Dec flat field correction to the colors relative to $r$ band from
stellar locus colors. 
We will take the resulting catalog (which
is calibrated to $1\%$ accuracy) as the truth and compare with our own 
measurements.
As \cite{Ivezic2007}  applied corrections that we did not, 
the comparison is not completely circular.

We sliced our star catalog into a series of 1 magnitude 
bins  and matched to the
\cite{Ivezic2007}  catalog using 
a $1"$ matching radius and discarding objects with more
than one match inside that radius. No flags were applied to
our star selection; in particular we did not demand an isolated,
well measured set of stars to start with. 

Table~\ref{tab:calcat} summarizes the results of our comparison.
We defined $\Delta_i$ as the median of the 
difference between our measurements
and the quantities reported in the standard star catalog, 
for  magnitudes ($i=m$) and colors ($i=c$). 
$\Delta_i$ is a measure of the zero-point offset.
Statistical uncertainty in the zero-points,
obtained as the rms of the differences,
are of the order a few millimag$/\sqrt{N_{\rm{obj}}}$; 
as usual in photometry we can expect systematics to dominate this.
The offsets from the standard zero-points are less than 
5 millimags ($1\%$ photometry corresponds to 10 millimags)
in all cases.

\begin{deluxetable*}{cccccccccc}
\tablecolumns{10}
\tablecaption{Relative Zero-points for Selected Coadd Catalog Samples}
\tablehead{\colhead{filter} & \colhead{$N_{\rm{obj}}$}%
         & \colhead{mag}    & \colhead{$\Delta_m$}%
         & \colhead{color}  & \colhead{$\Delta_c$}%
         & \colhead{$\langle\Delta_m\rangle_{\rm{RA}}$}    & \colhead{$\langle\Delta_c\rangle_{\rm{RA}}$} %
         & \colhead{$\langle\Delta_m\rangle_{\rm{Dec}}$}   & \colhead{$\langle\Delta_c\rangle_{\rm{Dec}}$}
}
\startdata
u & 1311 & 20-21 & $1.2 \pm 1.6$ & u-g  & $-4.3 \pm 3.6$ & $-21.0       \pm 8.2$  & $-12.1        \pm 8.1 $ & $-21.3        \pm 28.8$    & $-12.4  \pm 15.3$     \\
g & 1399 & 19-20 & $2.2 \pm 1.6$ & g-r  & $-1.7 \pm 1.4$ & $\phm{-2}6.5 \pm 4.6$  & \phm{1}$-1.2  \pm 3.5 $ & \phm{-11}$5.8 \pm 10.2$    & \phm{1}$-1.1 \pm \phn6.3$ \\
r & 3703 & 19-20 & $3.7 \pm 1.5$ & r-i  & $-2.9 \pm 2.2$ & $\phm{-2}3.2 \pm 2.5 $ & \phm{-11}$4.0 \pm 2.1 $ & \phm{-11}$3.2 \pm \phn6.3$ & \phm{-11}$4.0 \pm \phn4.1$ \\
i & 7436 & 19-20 & $0.9 \pm 2.0$ & i-z  & $-1.5 \pm 2.3$ & $\phm{-2}0.1 \pm 2.0 $ & \phm{-11}$6.3 \pm 3.1 $ & \phm{-11}$1.0 \pm \phn6.2$ & \phm{-11}$5.7 \pm \phn6.8$ \\
z & 3963 & 18-19 & $4.3 \pm 2.4$ & \nodata  &  \nodata   & \phm{2}$-5.7 \pm 3.3 $ & \nodata                 & \phm{1}$-5.3  \pm 11.7$    & \nodata 
\enddata
\label{tab:calcat}
\tablecomments{Rows correspond to samples created independently for each filter. Magnitude ranges are indicated by the mag column. 
$N_{\rm{obj}}$ is the number of stars in each sample. $\Delta_i$ is the median zero-point difference, in millimags, for either magnitudes ($i=m$) or colors ($i=c$) and 
$\langle\Delta_i\rangle_{j}$ is the median of the mean difference in spatial bins, RA or Dec.}
\end{deluxetable*}

\begin{figure*}
\centering
\includegraphics[width=1.0\linewidth]{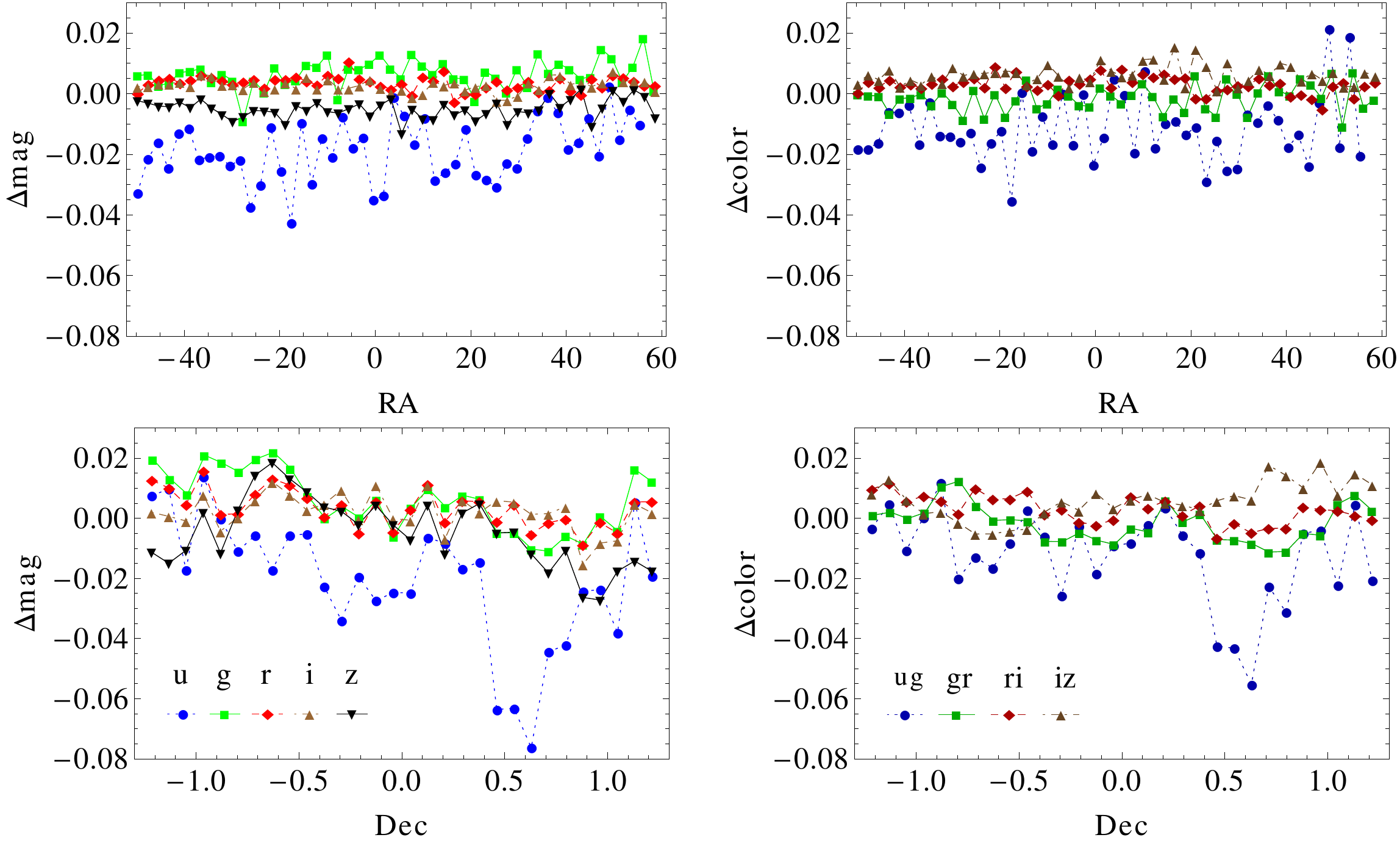}
\caption[Calibration residuals plot]{
Zero-point offsets measured in bins of RA and Dec,
showing spatial variations in the photometric calibration of 
the coadd data.
}
\label{fig-cal-plots}
\end{figure*}

We also examined the spatial variations of the zero-point
offset and its uncertainty. To this end, we first took 
50 equally spaced bins of width $2.2\arcdeg$ in RA, 
and computed the mean differences in magnitudes and 
colors for each bin. We repeated this procedure 
in Dec bins, choosing 30 bins of $5\arcmin$ width.  
Figure~\ref{fig-cal-plots} shows these
mean zero-point offsets as a function of 
RA and Dec, indicating that spatial variations 
non-negligible, specially for $u$ band. 
As a measure of the overall offsets we 
computed the median of those means,
$\langle\Delta_i\rangle_{j}$, where $j$ means
either RA or Dec. These values are also
included in Table~\ref{tab:calcat} with 
the uncertainties  estimated as the rms or 
the means.  
 
Our  comparison to the standard star catalog 
reveals, therefore, 
that along the RA axis in
Stripe 82 the calibration varies 5 millimag  
in $g,r,i,z$, \gr, \ri, \iz
$\le 10$ millimag  in $u$ and \ug. Likewise, 
along the Dec axis
the calibration varies by
$<5$ millimag in $g,r,i,z$, \gr, \ri, \iz
and 30 millimag for  $u$, 20 millimag  for \ug and $\le 10$ millimag for the others.
These can be seen in Fig.~\ref{fig-cal-plots} (see also Table~\ref{tab:calcat}).
The variation in Dec is significantly larger than that in RA,
probably reflecting systematic errors in the
 PT flat field images and thus in the 
PT standard overlap fields the used for the
calibration. 

\subsection{PSF Modeling}
\label{sec-psfModeling}

\begin{figure*}
\includegraphics[width=1.\linewidth,trim=110 0 110 0]{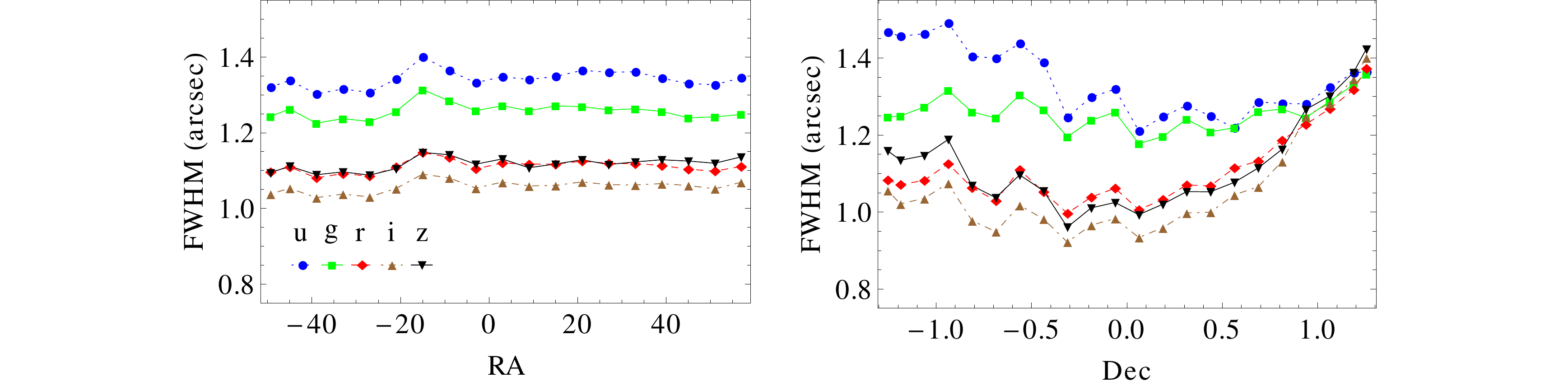}
\caption[Seeing in the coadd.]{
Left: Mean FWHM as a function of RA for the five filters.
The range of seeing values is consistent with the expected Kolomgorov $\lambda^{-0.2}$ 
scaling with the exception of $r$ and $i$ bands.
As the input images were selected in $r$ 
and data in the other bands are taken at a few minutes removed in time in the order
$riugz$, one might expect this behavior.
Right: Mean FWHM as a function of Dec. 
At Dec $> 0.5$ the seeing gets worse as the camera optics begins to dominate
(see, e.g., \cite{edr}).
}
\label{seeing-ra}
\end{figure*}

We verified how well we modeled the PSF in several ways, starting by
computing the spatial variations in seeing, which can be written as
\begin{equation}
{\rm FWHM(arcsec)} = 2 \sqrt{2\ln{2}}\sqrt{\rm{mrrcc}/2}\cdot S
\end{equation}
where $S=0.396\arcsec$ is the pixel scale and mrrcc is the sum
of the second moments of the PSF. 
We used high S/N stars in each bandpass for this test. Our results, 
illustrated on the left panel of Fig.~\ref{seeing-ra}, show that the 
seeing is best in the redder filters, consistently with a 
Kolmogorov seeing law with the exception of $r$-band and $i$-band. We  
interpret this as an effect of selecting the input frames 
for the coadd in $r$-band (see Table~\ref{crittable}) associated with an 
effect of the time scales of the Kolmogorov law.
Recall that the data making up a field are taken at different times, ranging over
8 minutes from $r$-band to $g$-band. Our data support the assumption that
the time scales of Kolmogorov seeing
are such that the seeing is uncorrelated after $\sim 1$ minute, but since
$i$ and $r$ are next to each other in the imager, they can be correlated.
The median of the seeing of the single pass images 
are a few tenths of an arcsecond worse
than the points on this plot; this reflects the weighting of the coadd
(Eq.~\ref{eq-weights}).

Figure~\ref{seeing-ra} also show
the seeing as a function of Dec, averaging over RA.
The seeing is affected by the camera optics, which causes the upturn at one
end of the camera, 
The N strip (run 206) has about 0.075\arcsec\ worse seeing in $ugiz$, 
and 0.15\arcsec\ worse seeing in $r$, then the S strip (run 106), presumably due to
the statistics of the seeing in the input images. However,
at Dec $\gtrsim +0.5\arcdeg$,  as the camera optics begin to dominate, 
the seeing difference becomes negligible.

\begin{figure}
\includegraphics[width=1.0\columnwidth]{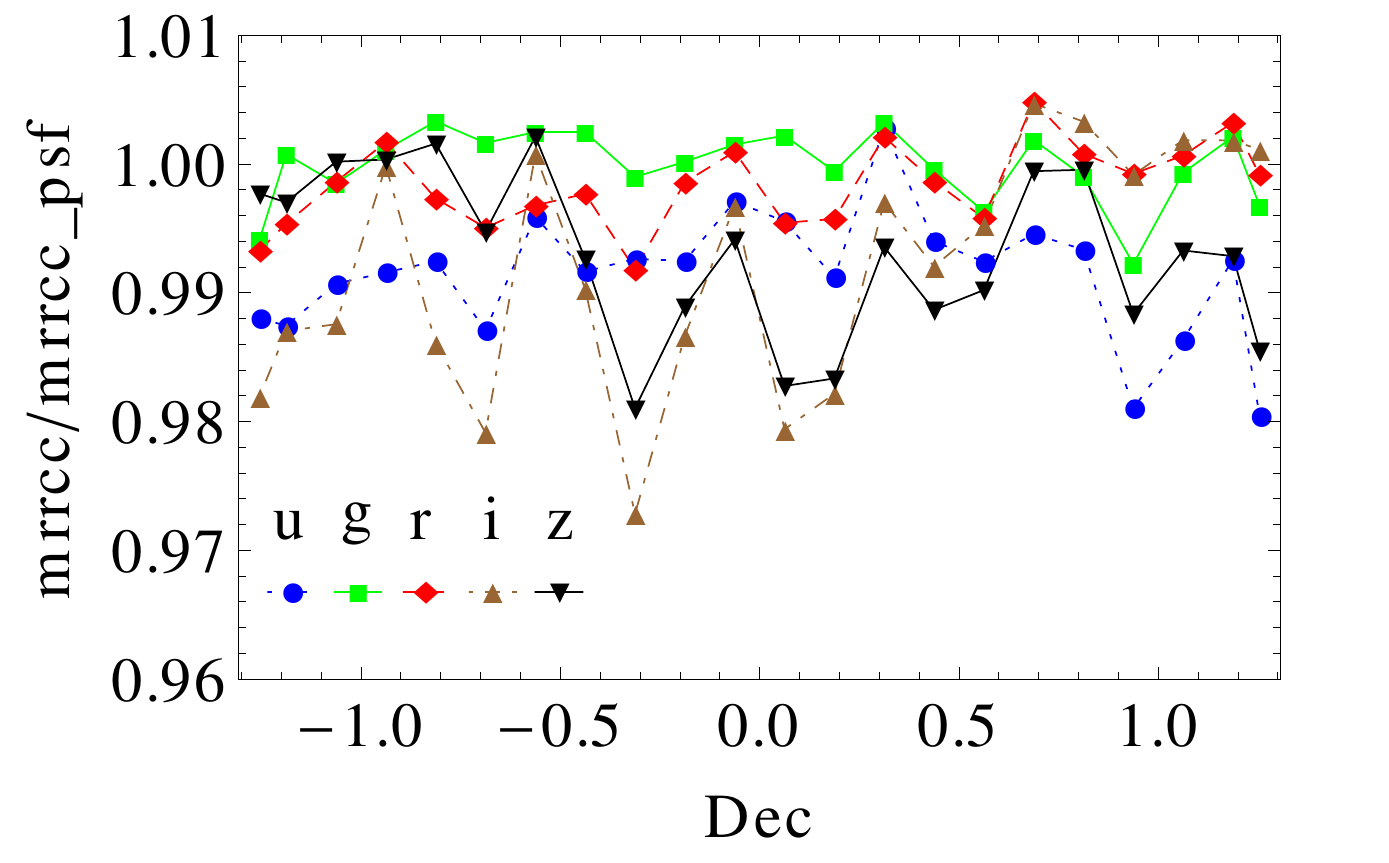}
\caption[Seeing in the coadd.]{ 
The declination projection of the ratio of size, measured through $m_rr_cc$,
of stars and the psf evaluated at the star positions 
({\tt mrrcc}/{\tt mrrcc\_psf}).
The ratio is accurately unity without dependence on declination, implying
that the PSF modeling is accurate.  }
\label{psf}
\end{figure}

Another interesting test of our PSF modeling is to check the 
reconstructed PSF at the
position of stars.  In Fig.~\ref{psf} we show the mean ratio of {\tt mrrcc} 
for stars and the reconstructed  PSF at the locations of stars, as a function
of Dec.  The strong declination
dependence seen in Fig.~\ref{seeing-ra} is not apparent. This suggests that
the modeling is fitting the spatial variation of the PSF well.
We  also found no correlation between the
statistics of star galaxy separation and
column number, again suggesting reasonable success in the PSF modeling. 

\begin{figure}
\includegraphics[width=1.0\columnwidth]{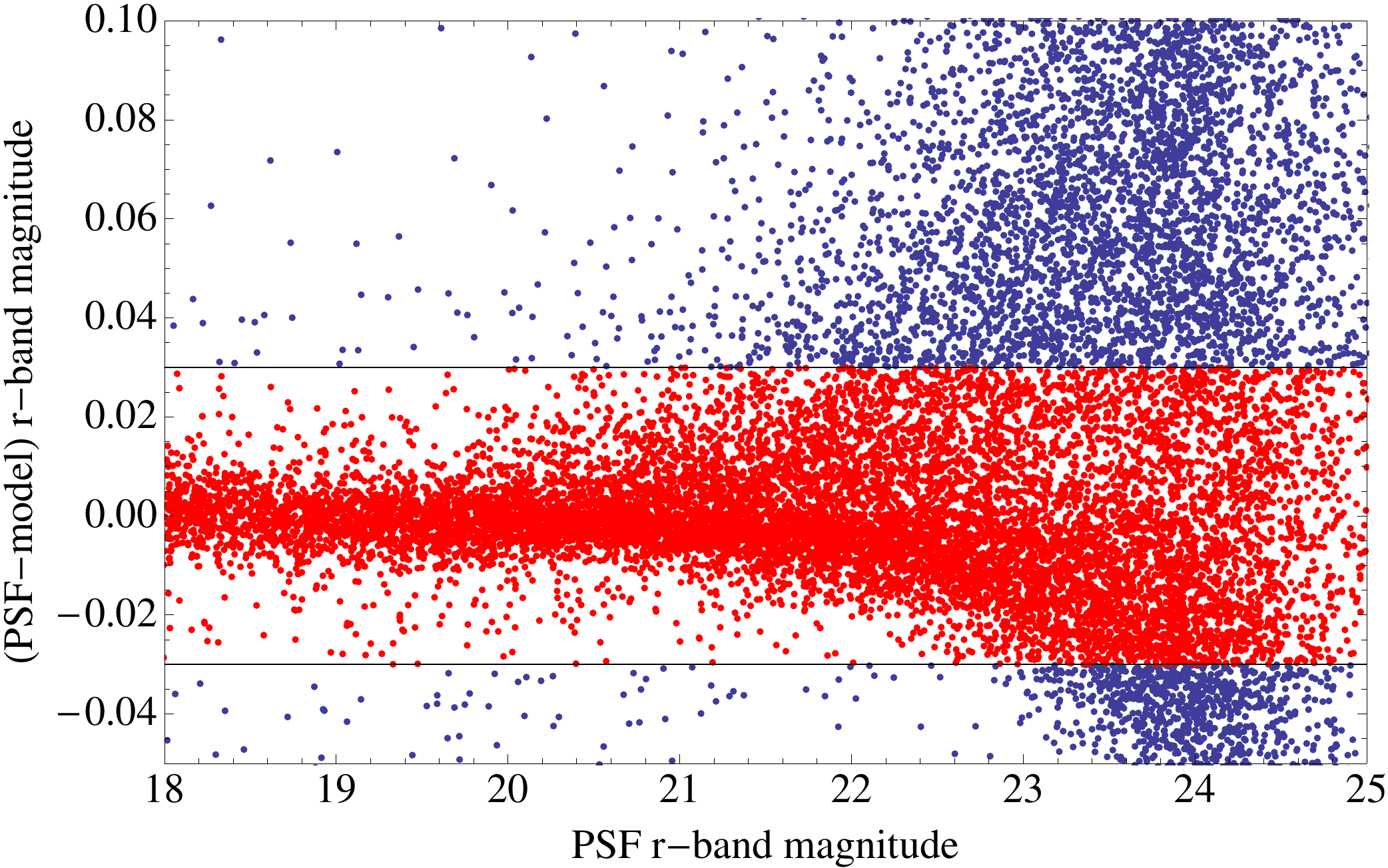}
\caption{
Star/galaxy separation based on PSF$-$model magnitude, 
$r_{\rm{psf}}-r_{\rm{model}}$. 
A model magnitude is the best galaxy deVaucouleurs or exponential profile
convolved with the PSF in the sense of describing the data. In the event
of a mismatch of the modeled PSF with the real PSF, or extended light,
the model magnitude will measure  more light than the PSF. It is thus
a sensitive indicator of whether an object is well fit by the
modeled PSF.
Objects classified as stars are marked in red, galaxies in  blue. 
There is a clear stellar locus at $r_{\rm{psf}}-r_{\rm{model}}\approx 0$.
That the stellar locus bends towards negative values at $r_{\rm{psf}} > 22$ is likely 
due to errors in the PSF fit. 
}
\label{star-gal-sep}
\end{figure}

However, given the importance of the PSF modeling for crucial aspects of 
the data, such as star/galaxy classification,
accurate photometry and shape measurements, we examine it in more detail. 
A sensitive test of the PSF modeling is the $r$-band ``PSF minus model'' plot, 
$r_{\rm{psf}}-r_{\rm{model}}$~vs.~$r_{\rm{psf}}$.
This is shown in 
Fig.~\ref{star-gal-sep} for both stars (red) and galaxies (blue)  
(for details on the star/galaxy separation, see \S\ref{stargalx}).
Stars are expected to be found at a very narrow region around 
$r_{\rm{psf}}-r_{\rm{model}}=0$.
and this is clearly the case for the bright magnitudes in our plot. 
There is a noticeable trend towards negative values of 
$r_{\rm{psf}}-r_{\rm{model}}=0$
at the faint end, around magnitude 23, and an
upward spread at about magnitude 22.  The upward spread cannot
be due to galaxies behind  stars causing the PSF to slightly
broaden as this was excluded by checking the spatial statistics. 
These two features in the diagram suggest that we have
magnitude dependent PSF fitting problems which may introduce systematics 
in analyses involving the coadd data. 

While probing this, we found that during the processing described in
section~\S\ref{sec:wcm} the coadd incorrectly set the 
error propogation value to $T$ instead of the correct value of $T^2$. 
This means that the weight actually used was:
\begin{equation}
w = {{1}\over{{\rm FWHM}^2 \sigma^2}}
\end{equation}
This has little effect on the coadded image as we only selected those 
images with  $1/T \le 1.2$. The PSF photometry, however, depends so
sensitively on the modeled PSF that this may be the root of the problem.
The PSFs were constructed using the correct weights (section~\S\ref{sec:psf}), 
but these differed from those of the images.
In an experiment we ran {\tt Photo}
with input weights that were pure inverse variance
instead of SN weight that the images were built with. 
The resulting psf-model vs psf plots
showed stars with deviations away from zero at 0.05 mags/mag level.
The effects we are seeing in Fig.~\ref{star-gal-sep} are at the
$\sim 0.01$mags/mag level, consistent with the lower affect the
tranmission has on the weights.
We conclude that the low level psf problems that the structure in 
Fig.~\ref{star-gal-sep}
indicates is due to this mismatch of weights between the coadd images and
the coadd photometry measurements.

\subsection{Galaxy Catalog Purity}
\label{sec-purity}

These misclassified stars are an issue as they lie in the magnitude-color space of LRGs.
The contamination level of the galaxy catalog by missclassified stars was 
estimated using the morphological-independent redshift survey catalog from 
the Virmos VLT Deep Survey \citep[VVDS,][]{vvds-deep},
in particular the the $i < 22.5$ 22hr field. 
We selected  the objects above 95\% confidence level,  which resulted in a 
catalog containing 5158 stars and 4264 galaxies. This catalog served as 
a truth table, to match galaxies in the coadd at $RA \sim -25\arcdeg$.
908 spectroscopic stars and 3438 spectroscopic galaxies matched galaxies in the 
coadd catalog, indicating a contamination level of 18\%.
This is $\sim 3 $ times higher than expected, assuming that the 
performance of the SDSS Star/Galaxy separator would be as good for the coadd as
for the single pass data. The field in question is among the lowest galactic
latitudes and highest stellar densities in the coadd, so this contamination rate
is approximately an upper limit. The actual contamination rate would 
be roughly proportional to RA in the coadd. 
 
We verified that most of the problematic objects are in a localized region of the 
psf$-$model vs. magnitude space, specifically inside  the triangular region 
\begin{eqnarray}\label{trianglecut}
\nonumber r_{\rm{psf}}-r_{\rm{model}} & < & 0.1 \cdot r_{\rm{psf}} - 2.1  \\
          r_{\rm{psf}}-r_{\rm{model}} & > & 0.03 \\ 
\nonumber            21 < r_{\rm{psf}} & < & 22.5 
\end{eqnarray}
This indicates that an improved star galaxy separator can be designed 
using morphology and presumably color cuts.

\subsection{Star \& Galaxy Catalog Completeness}

We used the package 2DPHOT \citep{2dphot} to measure the completeness of 
the coadd star and galaxy catalogs. 
This is done by adding simulated objects to the images and computing the 
recovery rate for both stars and 
galaxies. The parameters used to generate the objects are taken from the image itself.
2DPHOT first detects the objects in the image, performs star-galaxy separation and 
measures the photometric and structural (Sersic) parameters (\photo\
is not used in this process). Then it 
creates a list of objects that reproduces the magnitude and size distributions of the
stars and galaxies found in the image 
and adds these simulated objects to the image.
Finally, it measures the new image and 
computes the completeness as the
fraction of objects recovered in each magnitude bin. 

The resulting 2DPHOT completeness vs. magnitude curves, $C(m)$, are well fit by a   
Fermi-Dirac distribution function 
\begin{equation}
C(m) = \frac{f_0}{1+exp((m-\mu)/\sigma)}
\end{equation}
where $\mu$ is the magnitude limit of the catalog (defined to be the magnitude at which $C(m)=50\%)$,
$f_0$ is a normalization constant and the parameter $\sigma$ controls how fast the completeness falls when 
it reaches the completeness threshold.  We use this fitting function to determine the depth of the coadd 
galaxy and star catalogs, $\mu_G$ and $\mu_S$ respectively. 

\begin{figure*}
\centering
\includegraphics[width=0.8\columnwidth]{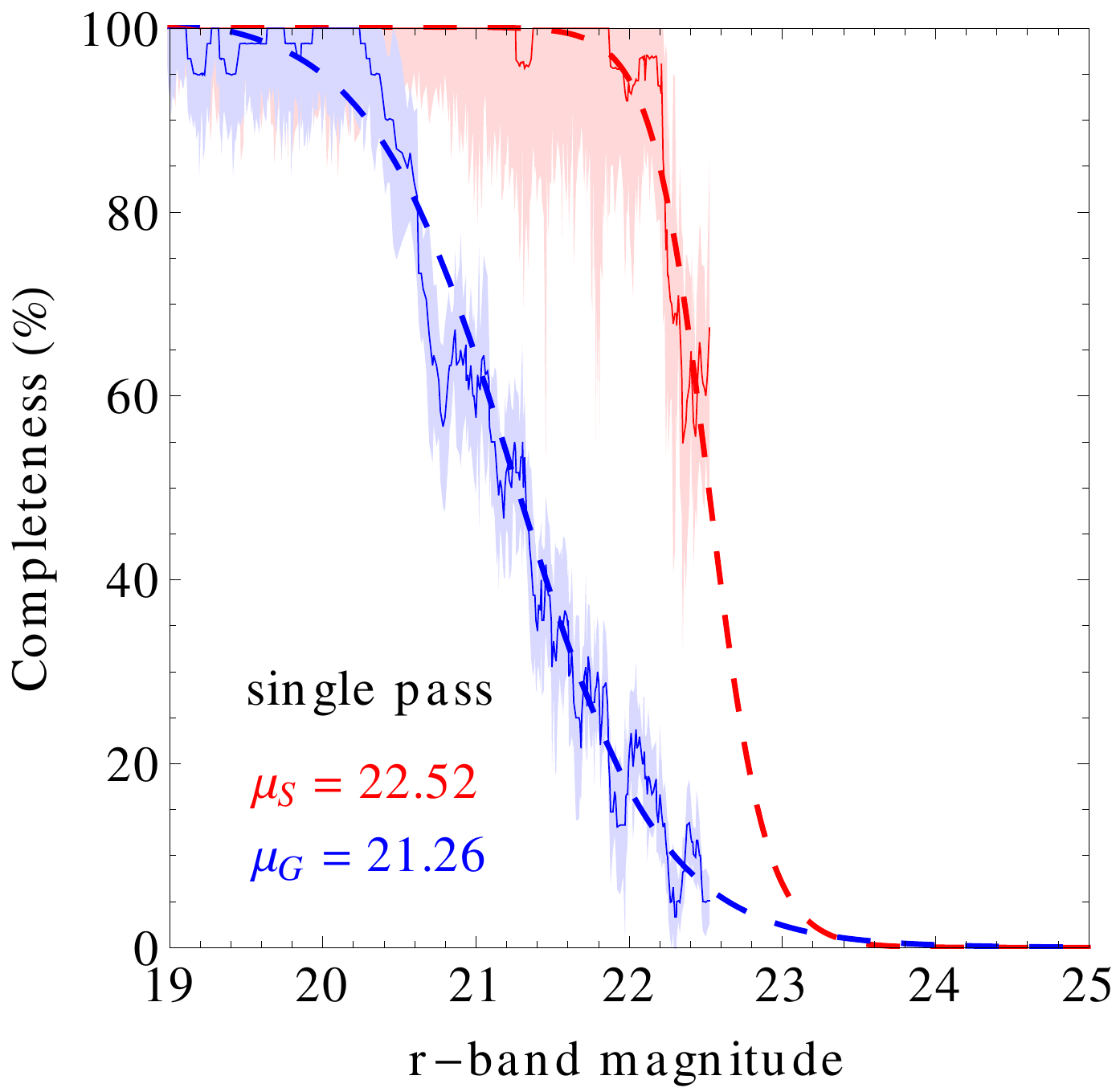}
\includegraphics[width=0.8\columnwidth]{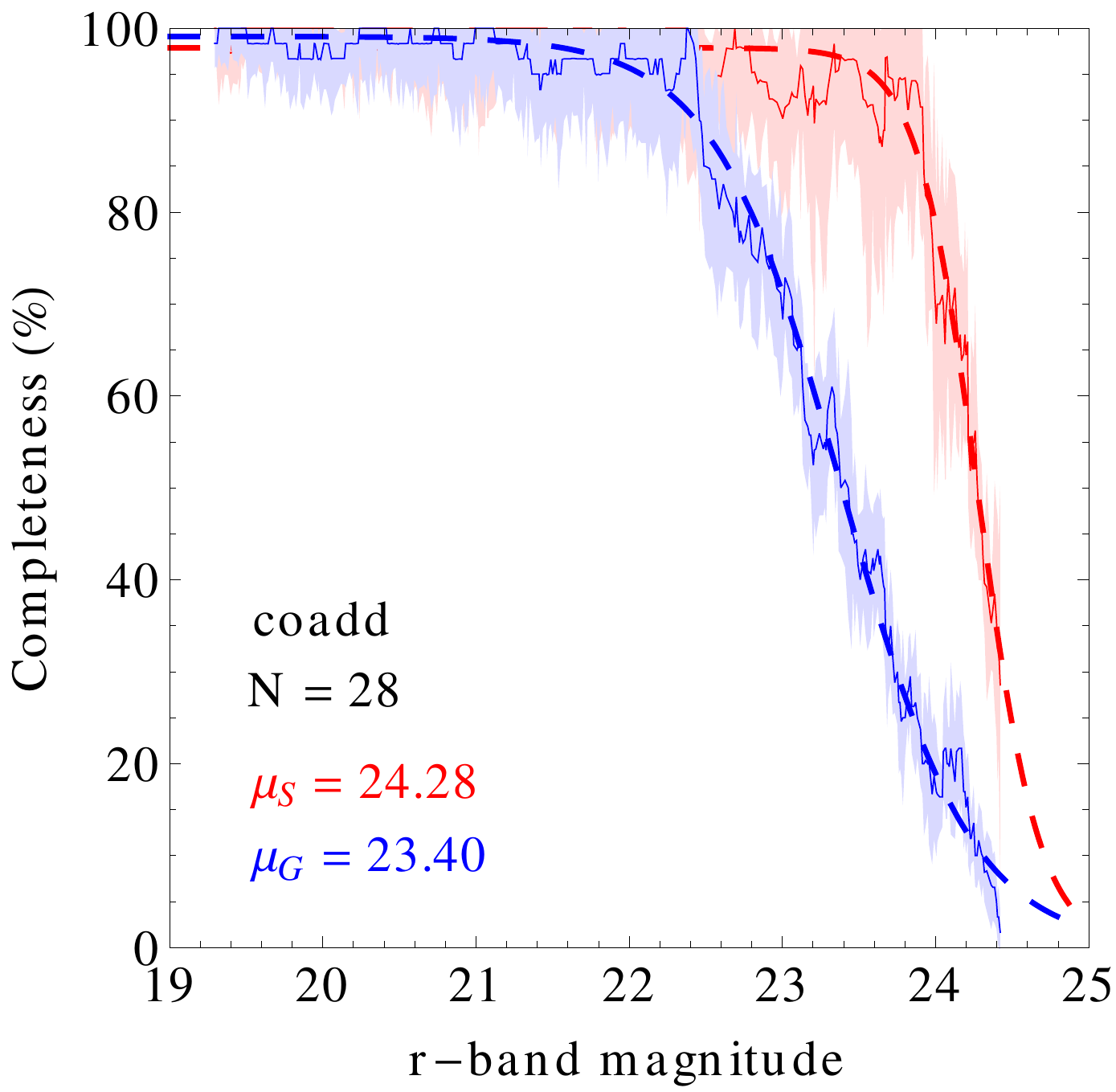} \\
\includegraphics[width=0.8\columnwidth]{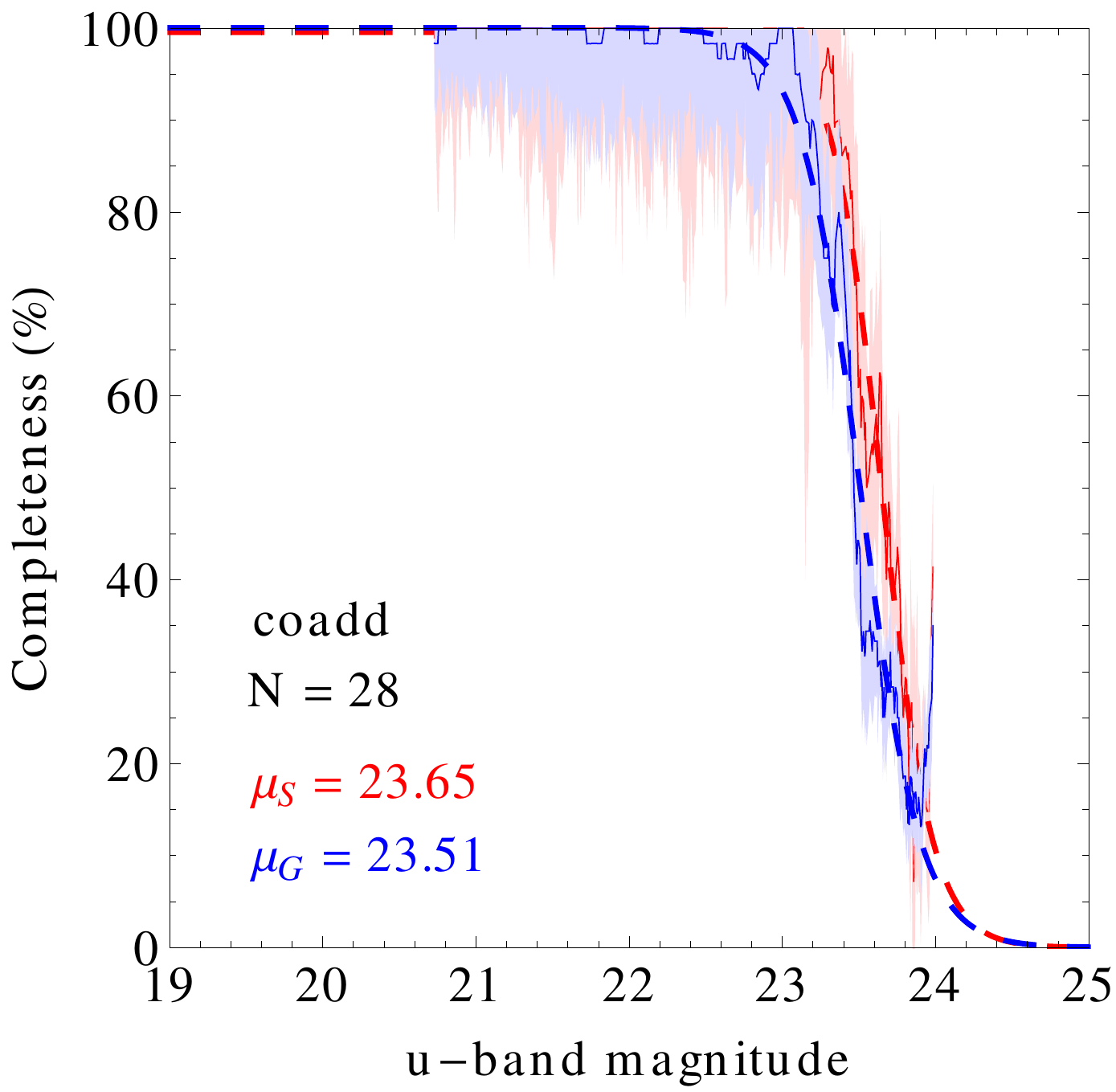}
\includegraphics[width=0.8\columnwidth]{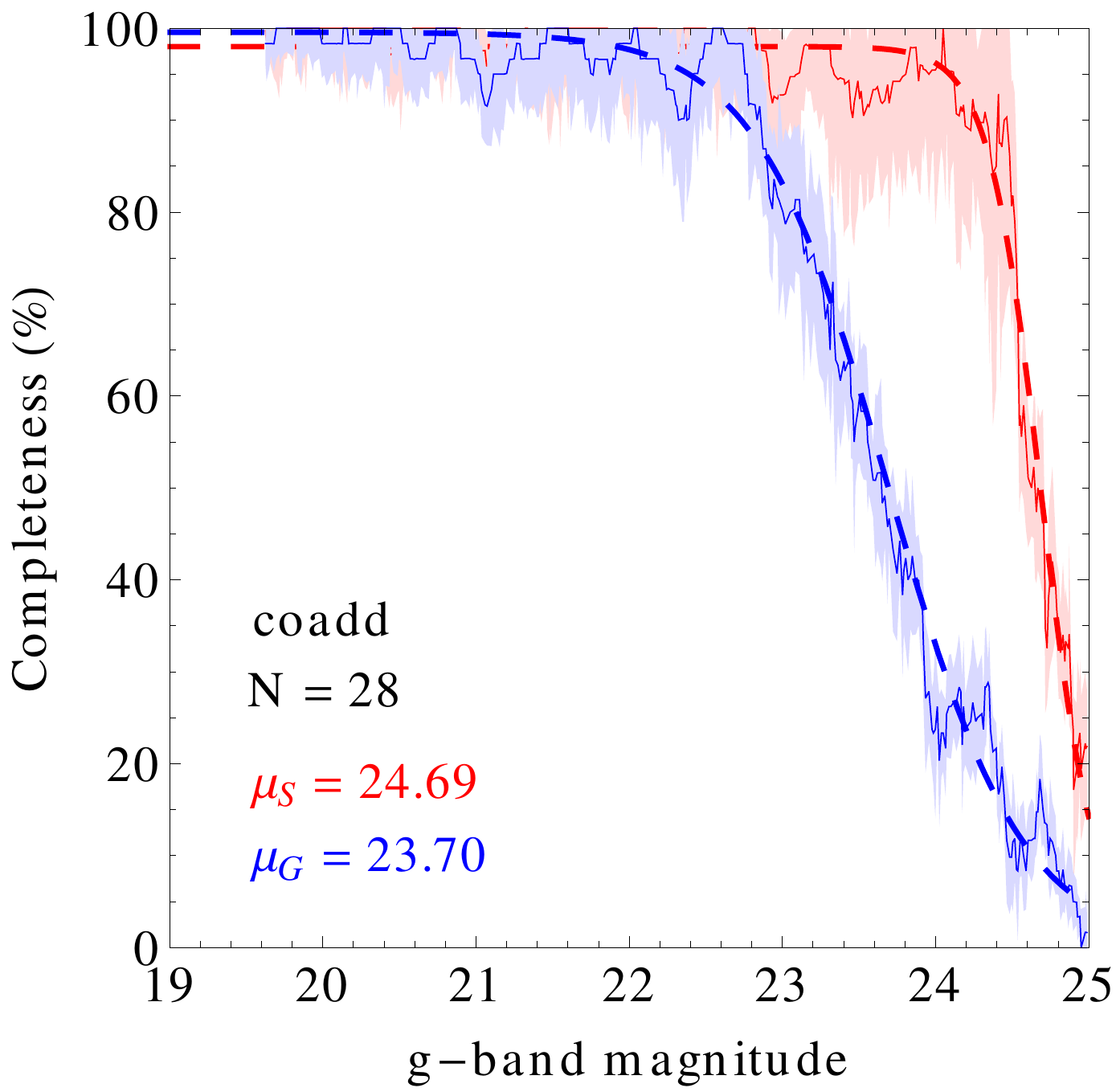} \\
\includegraphics[width=0.8\columnwidth]{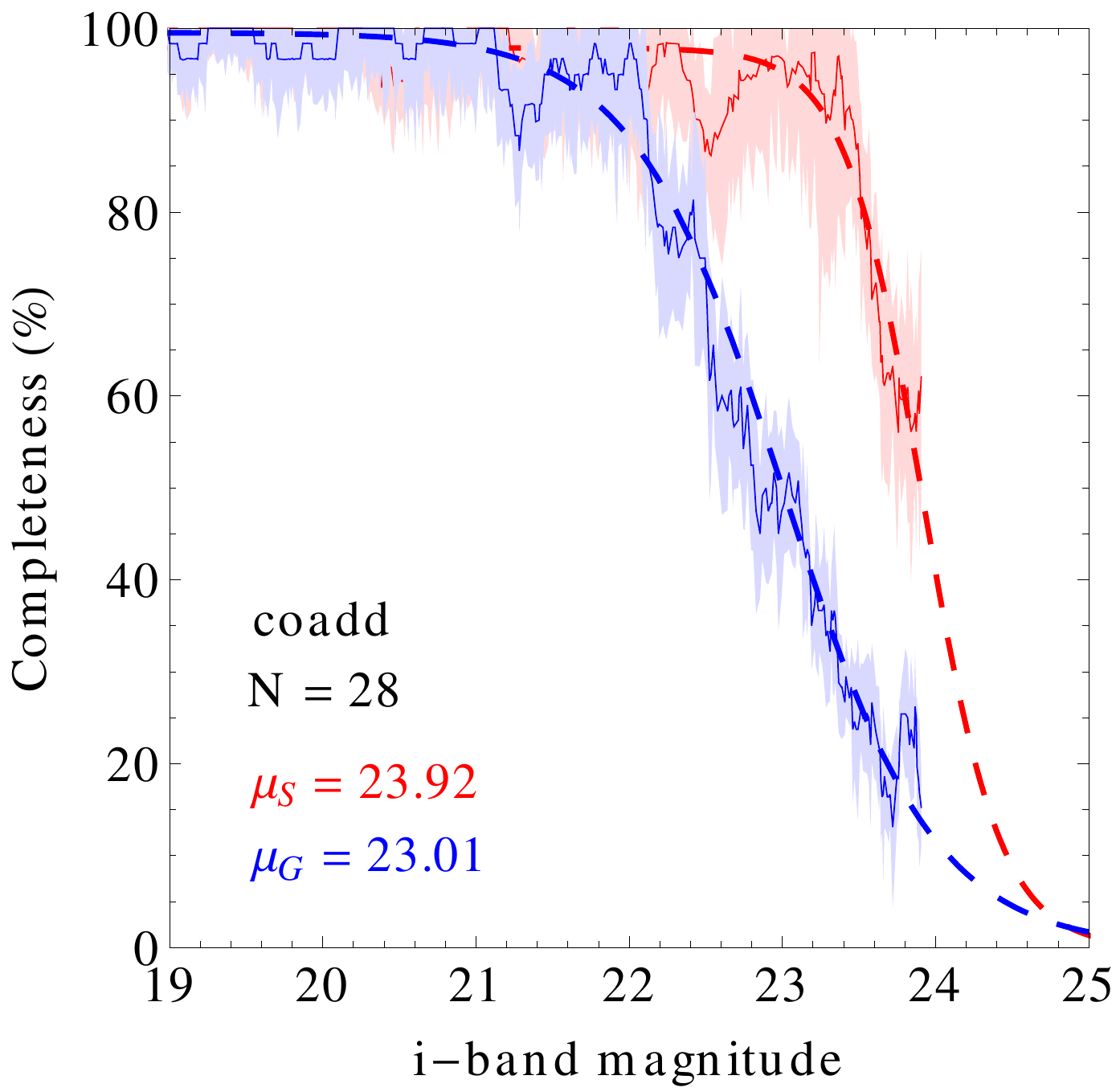}
\includegraphics[width=0.8\columnwidth]{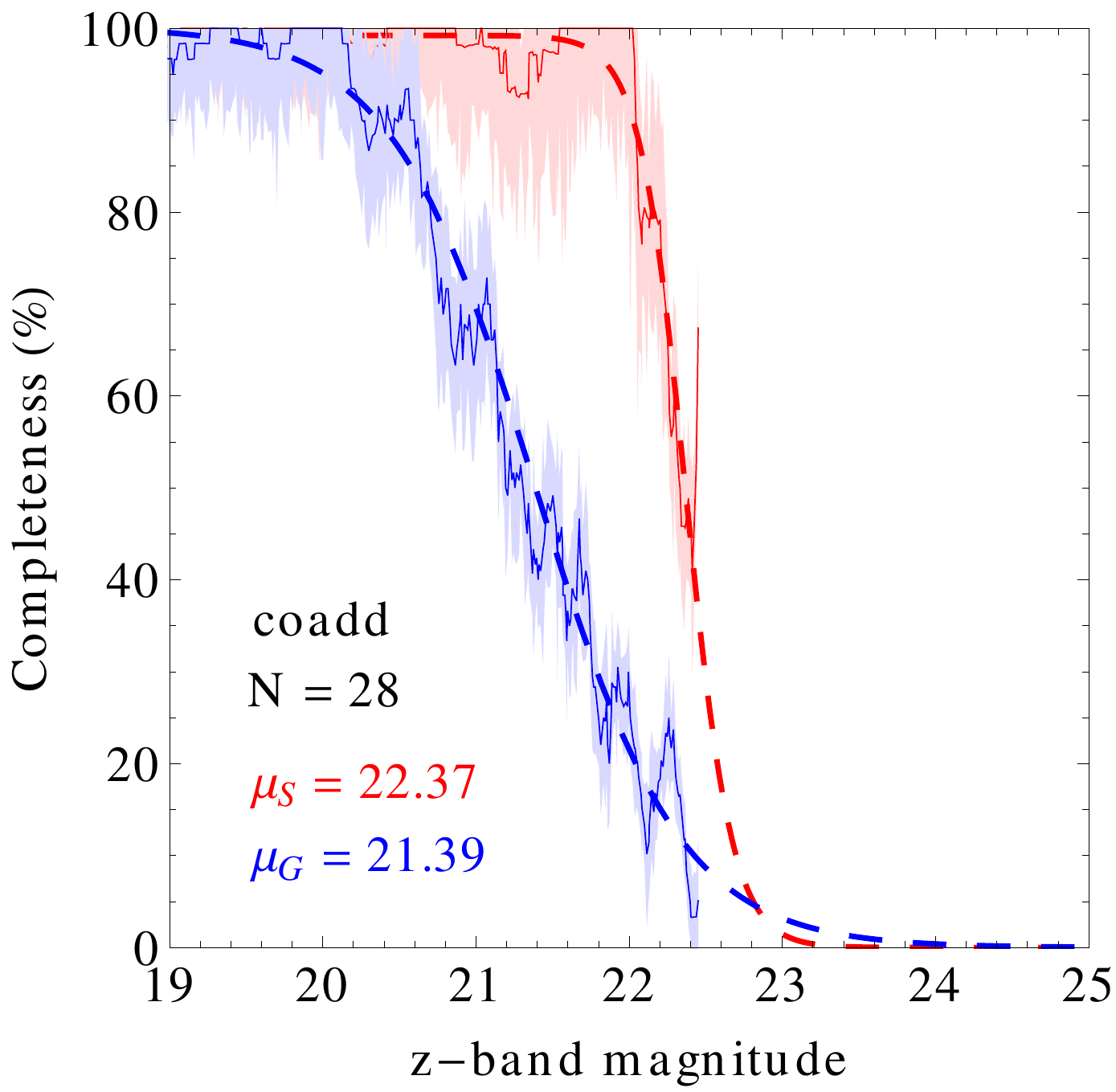}
\caption[Completeness curve for the Coadd galaxy catalog.]{
Completeness as function of $r$-band model magnitude for the
coadd  run 206, camcol 3, field 505 (corresponding to the 
image shown in Fig.~\ref{singleCoadd}). 
Point sources are represented in red, galaxies in blue. 
The solid lines are our measurements and the rms uncertainties 
are represented as light-colored regions. Dashed lines 
are the best fit model.   There is
a $\approx 0.09$ mag scatter in the 
measurement of the completeness level
as measured from frame to frame.
The numbers refer to the 
$50\%$ completeness level which we find
to be a much more stable fit than the $95\%$ level. 
Top: Comparison 
between the coadd (right) and one of the 
single pass images used in the coaddition (left),
showing that we achieve $\sim 2$ mag deeper as
$N=28$ images contributed into this particular frame.
Middle and bottom: Results for $griz$ on the same frame,
showing the typical depth of the coadd in each bandpass.
}
\label{82-2dphotresults}
\end{figure*}

Our results are illustrated in Fig.~\ref{82-2dphotresults}. 
The plots on the first row show the $r$-band completeness for
the same two fields pictured in Fig.~\ref{singleCoadd}. 
On the right we have the  coadd field 
(run 206, camcol 3, field 505) and for comparison, on the left, 
one of the 28 single pass images 
used as input for that particular field. 
The coadd reaches $r=24.3$ for point sources and $r=23.4$ for 
galaxies, going about 2 magnitudes deeper than a single pass 
image, as expected.  
The plots on the second and third rows show the coadd results for the other filters. 
Table \ref{tab:comp_lim} is a compilation of the 
median and rms values of the coadd magnitude limits, calculated for 50
fields randomly selected.  

\begin{deluxetable}{ccccc}
\tablecolumns{5}
\tablecaption{$50\%$ Completeness Limits in the Coadd}
\tablehead{filter  & $\mu_S$ & $\sigma(\mu_S)$ &  $\mu_G$ & $\sigma(\mu_G)$ }
\startdata
u & 23.63  & 0.06 & 23.25 & 0.23  \\
g & 24.56  & 0.10 & 23.51 & 0.18  \\
r & 24.23 & 0.08 &  23.26 & 0.14  \\
i & 23.74 & 0.15 &  22.69 & 0.17  \\
z & 22.29 & 0.09 &  21.27 & \phn0.23  
\enddata
\label{tab:comp_lim}
\tablecomments{Coadd magnitude limits for stars ($\mu_S$) and galaxies
($\mu_G$). Values reported are medians and rms  
calculated for 50 fields randomly selected accross Stripe 82.}
\end{deluxetable}

\subsection{Color-Color Diagrams}

\begin{figure*}
\centering
\includegraphics[width=1.45\columnwidth]{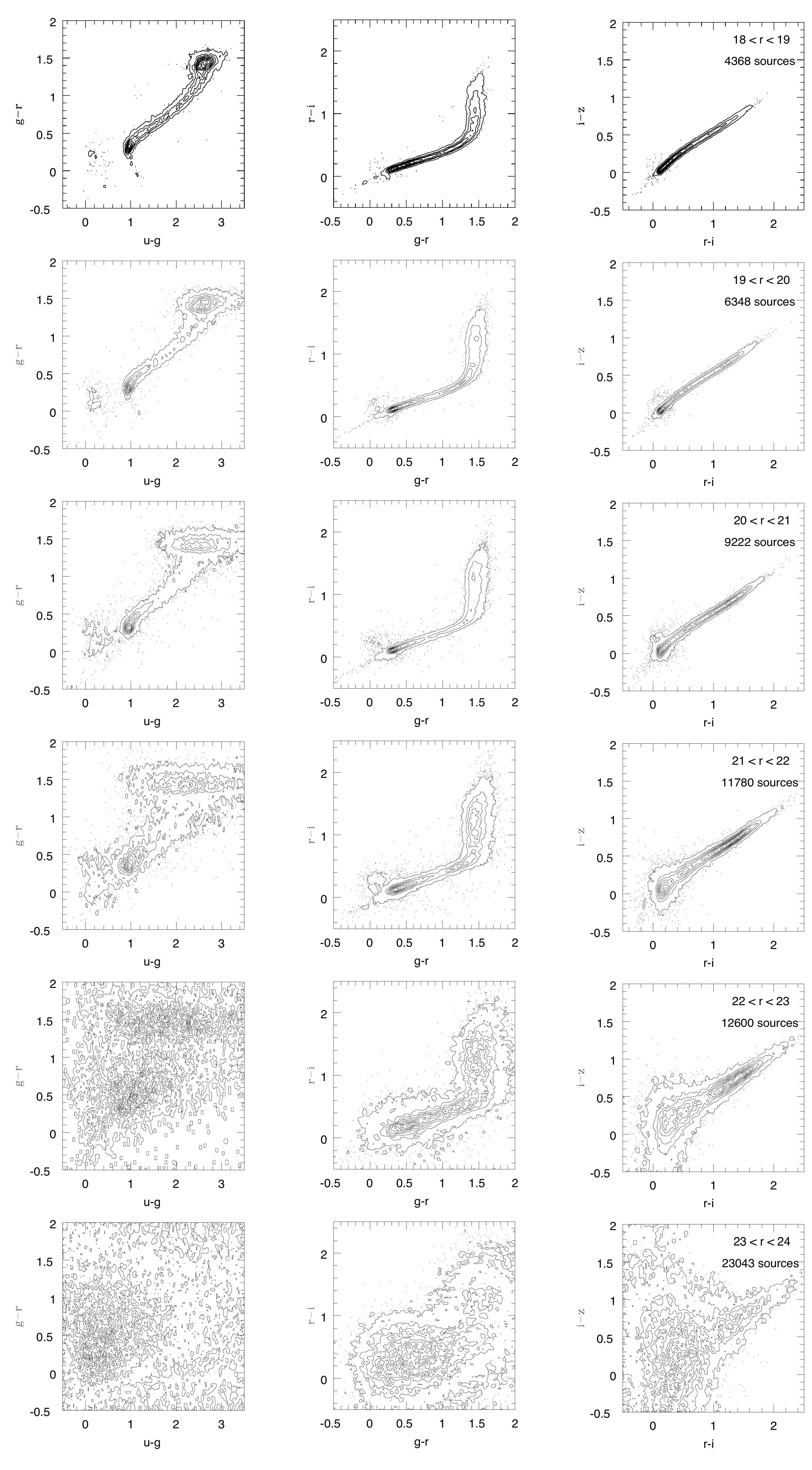}
\caption[color color diagrams.]{ 
Color-color diagrams of stars in a  12 degree$^2$ field, in bins of 1 magnitude
from $18 \le r \le 24$. Photon noise begins to dominate the bluer color
in the \ug\ vs \gr\ plot at
$r \ge 20$, the \gr\ vs \ri\ plot at $r \ge 21$, and the $z$ band in the
\iz\ vs $r-i$ plot at $r > 21$. As the stellar locus is intriniscly very
thin, these plots provide a sense of how good the photometry is at brighter
magnitudes, and where statistical noise begins to dominate.
}
\label{cc-panels}
\end{figure*}

Stars populate a well-defined locus in the color-color space almost
independent of magnitude. 
Therefore color-color 
diagrams are useful to access the quality of the photometry.
Figure \ref{cc-panels} shows the color-color diagrams of a 
isolated and well measured
sample of coadd stars 
selected at $-5\arcdeg \le$ RA $\le 0\arcdeg$, in a high galactic latitude in Stripe 82 (see query in the Appendix).
The sample was split into 1 magnitude bins in r-band. At brighter
magnitudes the intrisic thinness of the stellar locus is apparent.
At fainter magnitudes statistical noise begins to dominate. Since this is
an $r$-band sample, the $u$-band stars in the $20 < r < 21$ panel are quite
faint; one cannot read from these diagrams where the signal to noise of the
data degrades. One can read in these diagrams just how good the
photometry can be for clean samples of stars.

For a reasons which we have not unearthed, the satureted flag did not propogate through
\photo. Beware of objects with $m < 15.5$.


\subsection{Number Counts}
\label{sec-numberCounts}

\begin{figure}
\includegraphics[width=0.8\columnwidth,angle=-90,trim=0 40 100 0]{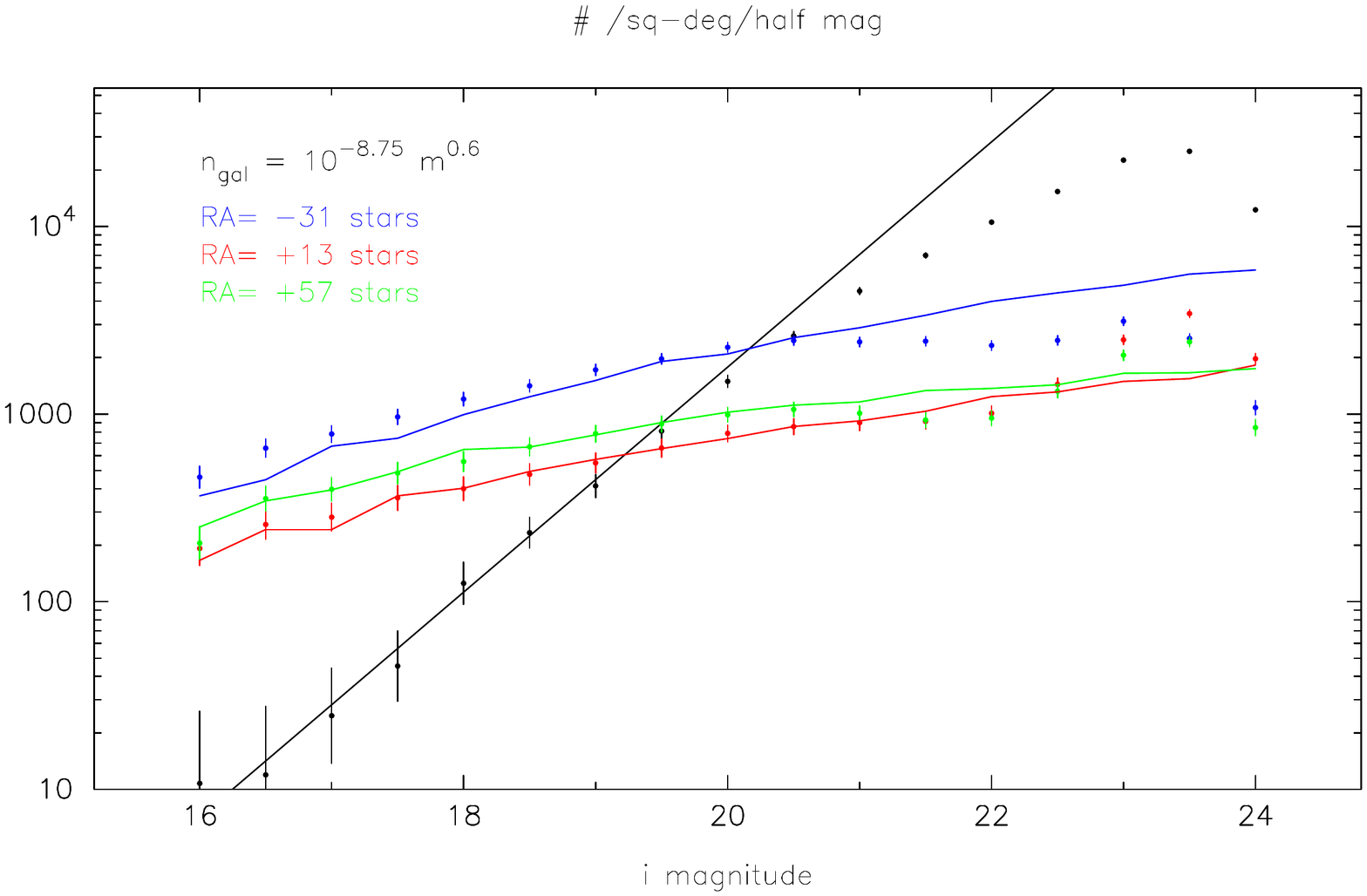}
\caption[Number Counts in the Coadd.]{
Star and galaxy counts in $5$deg$^2$ patches: the data are points with poisson errors,
and models are shown by lines.
The three patches are at
RA $= -31\arcdeg$, Dec $= 0\arcdeg$ ($l=58\arcdeg$, $b = -40\arcdeg$),
RA $= 13\arcdeg$, Dec $= 0\arcdeg$ ($l=123\arcdeg$, $b = -63\arcdeg$), and 
RA $= 57\arcdeg$, Dec $= 0\arcdeg$ ($l=188\arcdeg$, $b = -40\arcdeg$).
The galaxy counts (in black)
show the expected Euclidean power law at $i < 20$, the slow change of slope
due to cosmological volume and galaxy evolution at $20 < i < 23.5$, and
the roll off at $i \approx 23.5$ due to completeness issues.
The star count models
fit the data well enough for the desired purpose, until
$i \approx 23.5$, where there is a sudden upturn in star counts, most evident in the
RA $= 57\arcdeg$ data. This is certainly due to galaxies  being classified as stars.
}
\label{star-gal-counts}
\end{figure}

The number counts of stars and galaxies allow us to asses the depth of the 
coadd and the success of the star/galaxy separation. 
Figure~\ref{star-gal-counts} shows the $i$-band star and galaxy counts in 
regions of 5 deg$^2$ patches
at a variety of galactic longitudes and latitudes along Stripe 82.
The galaxy number counts 
show the Euclidean $m^{0.6}$ power law expected at $i < 20$, the slow change of slope
due to cosmological volume and galaxy evolution at fainter magnitudes, and a
roll off at $i \approx 23.5$ due to completeness issues. We have seen
in panel 5 of Fig.~\ref{82-2dphotresults} that the galaxy catalog is
 $\approx  50\%$ complete at $i=23$, consistent with the deviation from the
slowly rolling power law index seen here. We conclude that this plot
shows nothing seriously wrong with the galaxy counts from $16 \le i \le 23.5$.

The star counts are going to be slightly more problematic.
They are rough power laws that cross the galaxy counts at $i\approx 20$.
The expected roll-over from incompleteness is apparent near $i \approx 23.5$.
The steep rise in counts of stars seen easiest in the RA = $+57\arcdeg$ patch
data points is from galaxies being misclassified as stars: the galaxies
outnumber the stars by $\gtrsim 30$ so even a small misclassification rate
result in a large number of galaxies scattered into 
the star class, and one can see in Fig.~\ref{star-gal-sep} that at faint magnitudes
the galaxy locus crosses the stellar locus. Once it does, stars can no longer
be distinguished by galaxies using psf-model alone.

\begin{figure}
\includegraphics[width=1.0\columnwidth]{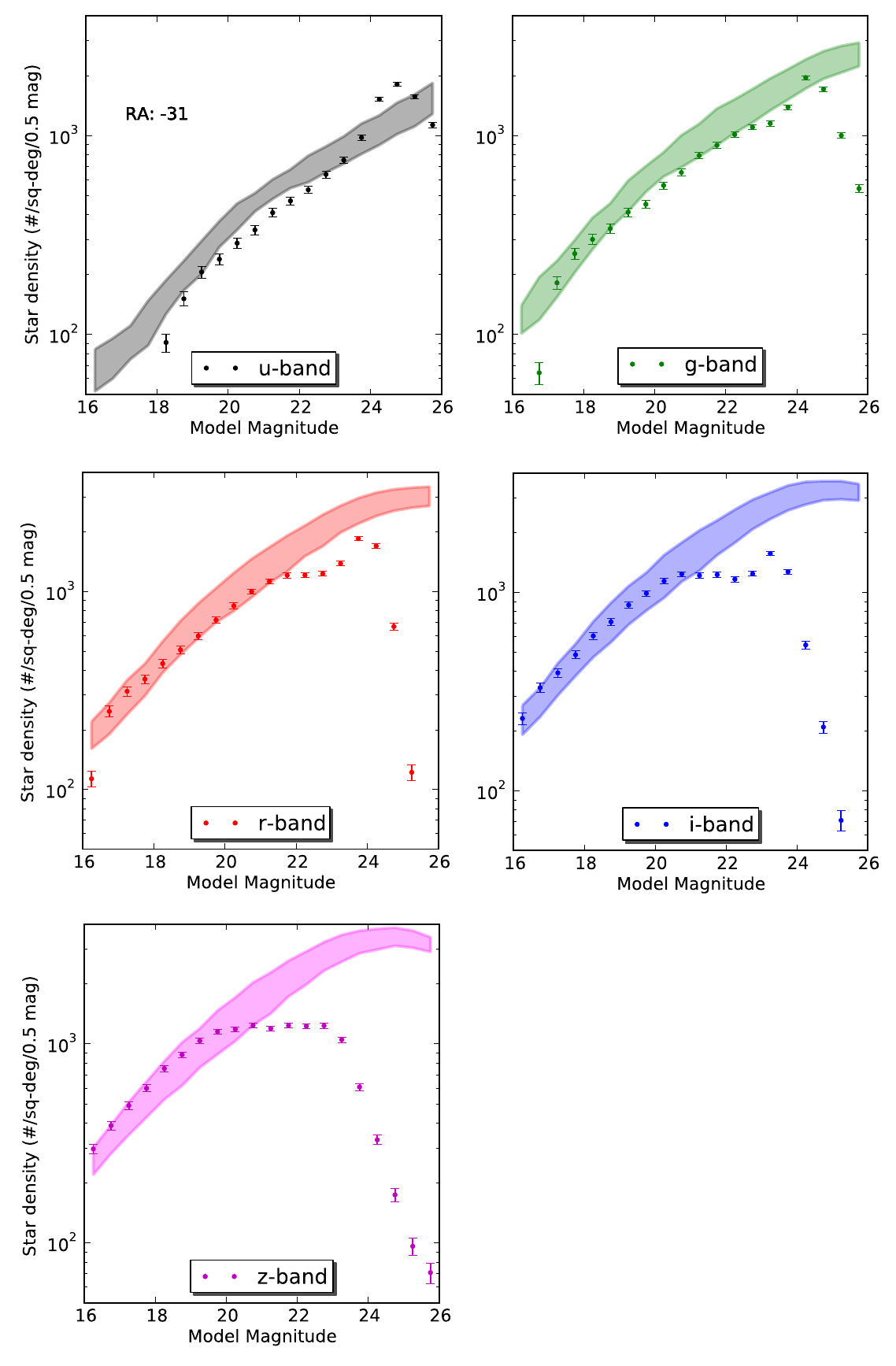}
\caption[Number Counts in the Coadd.]{
The Trilegal model star counts in u,g,r,i,z and coadd star counts at 
RA $= -31\arcdeg$, Dec $= 0\arcdeg$ ($b = -40\arcdeg, l = 58\arcdeg$).
The bands show a reasonable range of models, from one derived from Bochanski, 
the other from a private communication model of Santiago, 
modified to have the thin disk scale height $z_o = 80$ pc.
}
\label{star-counts}
\end{figure}

The models are from Trilegal star count modelling 
\citep[][online version v1.4]{trilegal}.
In Figure~\ref{star-gal-counts} the galactic model parameters are those
found by \cite{juric2008} using their SDSS star count tomographic mapping
technique. The slope of the rough power-law is due to a combination of
thin disk at brighter magnitudes and halo stars at fainter magnitudes.
This model doesn't fit the data particularly well at the RA $=-31\arcdeg$ 
patch,
or at lower RA. Models with a lower exponential scale height, such as
those derived from SDSS data using Trilegal model fits by B. Santiago 
(private communication) or from the SDSS m-dwarf fits of \cite{bochanski2007}
do describe the Stripe 82 data at RA $\lesssim -31\arcdeg$.
Figure~\ref{star-counts}  shows number counts in all five filters
for the RA = $-31\arcdeg$ field along with bands showing the range
of star counts from the Santiago and Bochanski galactic models.
This figure also shows, in all filters, the spike in counts as galaxies
enter the star catalog in large numbers at faint magnitudes, and
often a flattening of star counts at mag $\sim 22$. As we have
seen in section~\S\ref{sec-purity} stars are entering the galaxy catalog
in higher than expected numbers at these magnitudes due, most likely,
to a PSF modelling problem, and the flattening is probably a reflection 
of this. Nevertheless, the data seems to indicate that there is an 
interesting problem
in combining the deep Stripe 82 star counts with the single pass
SDSS star counts. Overall, this and the previous figure show reasonable 
agreement between the Stripe 82 star counts and the models. 
Comparison with star count models in this paper is an approximate
attempt to validate the overall sanity of the stellar counts. For a 
a detailed Galaxy study, see e.g.~\cite{Sesar:2010}.  


%
%
%



\section{Science}\label{science}
The coadd galaxy catalog is the largest homogeneous photometric
catalog of its depth. It has 13 million galaxies and we have shown
that it is complete to $r=23.5$, being 2 magnitudes deeper
than the SDSS single scan imaging survey. This implies that,
despite covering an area 40 times smaller,
the coadd volume has roughly 1/3 of the total
volume of the SDSS data on the Northern Galactic Cap.
The scientific questions that can be addressed by exploring
this wealth of data are numerous and diverse. In this section we
present a brief overview of recent and ongoing analyses using
the coadd data.

The presence of deep imaging provides a template for variable object
measurements and a means to check on the fainter objects
of the single scan SDSS data. The SDSS-II Supernova Survey \citep{sdssSN} used the 
coadd images as a high-quality, photometrically calibrated template for 
carrying out image subtraction to discover supernovae.
Later observations of the host galaxy often start with the coadd images
\citep{lampeitl2010,gupta2011} to locate the host.
As much of the Stripe 82 data was taken as part of the SDSS-II SN survey,
this use is expected.

There are many uses that are less tightly related to the goals of taking the data.
For example, \cite{liu2011} use the greater depth of the coadd as a means
to check their science at the main survey limits, as they  search for paired
quasars in the SDSS and use the coadd to check for the fraction
of interacting pairs missed due to the surface brightness limit of
the single scan.
\cite{jiang2008} used the coadd to perform the 
the discovery of 5 quasars at redshift $\sim 6 $ and
to compute the high redshift quasar luminosity function.
\cite{Vidrih:2007} find white dwarf candidates.
Our exploration of the Trilegal models in
section~\S\ref{sec-numberCounts} shows that these
data could be used to constrain Galactic models.

Our main interest is in cosmology, and here there is a clear
path of work.
Photometric redshift measurements for the coadd galaxies were
obtained by our group \citep{reis} using a neural network
trained on
spectroscopic redshifts obtained on Stripe 82.
As Stripe 82 is easily accesible by telescopes in both the
North and South hemispheres, the area has been well studied
spectroscopically. For the \cite{reis} work, we used the
Canadian Network for Observational Cosmology Field Galaxy Survey
\citep[CNOC;][]{Yee:2000}, the
Deep Extragalactic Evolutionary Probe
\citep[DEEP2;][]{Weiner:2005}, the
WiggleZ Dark Energy Survey \citep{Drinkwater:2010}, and the
VIsible imaging Multi-Object Spectrograph
Very Large Telescope Deep Survey \citep[VVDS;][]{vvds-deep}.
The mean photo-z error acheived was $\sigma(z)=0.031$ and the photo-z
catalog is  reliable out to $z\sim 0.8$, after which the
low S/N of objects in the SDSS $z$-band becomes
the limiting factor \citep[see;][for a detailed discussion]{reis}.
This photo-z catalog has been made public at the same time as
this paper as a value-added catalog.

Since the photometric redshifts go to $z\approx 0.75$, an
interesting program is
cluster finding in the range $0.5 < z < 0.75$.  Preliminary
cluster catalogs have been pursued by our group using the
Gaussian Mixture Brightest Cluster Galaxy
\citep[GMBCG;][]{hao2010} and the Voronoi Tessellation cluster
finder in 2+1 dimensions \citep[VTT]{vtt} algorithms.
We plan on pursing  a search for blue clusters
using the VTT as a finder and the GMMBCG
as a red sequence measuring engine aiming at studying
cluster formation and evolution.

In \cite{lin2011} we report the measurement of cosmological
parameters from the cosmic shear signal in the coadd.
All weak lensing analyses require accurate shape estimation
parameters, but the cosmic shear is an extreme case, due to its
very low signal.  \photo\ measures second moments and
related parameters needed for weak lensing, but
several systematic errors in the PSF had to be corrected.
Some of these systematic were doubt due to our mis-modelling
of the PSF because of the mis-matched weights (see section~\S\ref{sec-psfModeling}).
After the corrections and weak lensing-specific
quality cuts, the coadd data provides
$\sim 6$ galaxies per arcmin$^2$ for the analysis,
6 times more than the SDSS single pass data.
\cite{huff2011}, in an independent work, have also
measured the cosmic shear on Stripe 82. They did not use
the coadd described in this paper, but instead, made their own
coadd optimizing for weak lensing.

A related program is to measure the masses of clusters found in
the SDSS Coadd area. The MaxBCG \citep{maxbcg} cluster catalog
overlaps with Stripe 82 using single pass data
(as do other cluster catalogs, e.g.,
\cite{Geach:2011,dong2011,szabo2011}).

We have performed a stacked cluster weak lensing
analysis with the MaxBCG clusters as lenses and the coadd
galaxies as sources \citep{simet}. We divide our cluster
sample in bins of richness and measure a mass-richness
relation consistent with previous work \citep{Johnston:2007}.
This demonstrates that the coaddititon process does not dilute
the lensing signal.
As we detect an increasing  signal as a function of
source redshift we also conclude in \cite{simet} that we have
detected weak lensing tomography signal in the coadd.

Since the survey covers a large area on a part of the
sky that has been heavily
studied there are also many opportunities for multi-wavelength studies,
from the x-ray \citep[XMM:][]{xcs} to the microwave 
\citep[SZ:][]{menanteau2010,hand2011,reese2011,sehgal2011}.

It is unlikely that we have surveyed all of the science the
coadd has been put to use for, as it is part of the DR7 database
and easily accesible by the community. This paper serves as
as a technical description of this widely available dataset.

\section{Conclusions}\label{conclusion}

The SDSS performed repeat scanning of the equatorial region in the
South Galactic Cap known as Stripe 82. The amount of data observed
was comparable to the single scan coverage of the SDSS footprint.
The aim of the work described here was to coadd this data and
analyze it using the SDSS pipeline framework, notably the \photo\ pipeline.
Roughly a third of the existing Stripe 82 data was coadded, limited in time
by when the work was performed. The runs included calibrated and uncalibrated
data, so a relative calibration scheme was developed and applied. 
The images were mapped onto a SDSS run format output grid using the SDSS
astrometry, were coadded using a S/N weighting that includes seeing, and
inverse variance maps computed. The coadded images were run through
\photo\ using PSF models computed by coadding the PSFs.

The resulting catalogs have median seeing in $r$ of $1.1\arcsec$ and varies band
to band following Kolmogorov scalings. 
The catalogs are $50\%$ complete to $r=23.5$ (galaxies) and $r=24.3$ (stars). 
The photometry is good to $0.5\%$ in g,r,i, and
$1\%$ in u and z, as measured against the \cite{Ivezic2007} star catalog.
The PSF is modeled and despite minor issues it is useful for precision photometry.
Color-color diagrams of stars show a sharp and thin stellar locus, and
the number counts of stars at a variety of positions show agreement with
reasonable galactic models.
There are identifiable regions in psf-model vs magnitude space where
stars are being misclassified as galaxies, and we give suggestions on
how to eliminate these. Thus we have constructed high quality 
SDSS catalogs on the Stripe 82 region from images that are 
two magnitudes deeper than the single pass SDSS data.

The coadd catalog is largest homogeneous precision photometric catalog 
complete to r=23.5. 
The catalog has 13 million galaxies, and has a variety of uses from galactic
structure to large scale structure and weak lensing, to cosmology.
The data, including both images and catalogs,  are  available 
through the standard  SDSS distribution channels.

\acknowledgments
This research was done using resources provided by the Open Science Grid, 
which is supported by the National Science Foundation and the 
U.S. Department of Energy Office of Science.

Funding for the SDSS and SDSS-II has been provided by the Alfred P. Sloan Foundation, 
the Participating Institutions, the National Science Foundation, 
the U.S. Department of Energy, the National Aeronautics and Space Administration, 
the Japanese Monbukagakusho, the Max Planck Society, 
and the Higher Education Funding Council for England. 
The SDSS Web Site is http://www.sdss.org/.

The SDSS is managed by the Astrophysical Research Consortium for the 
Participating Institutions. The Participating Institutions are the 
American Museum of Natural History, Astrophysical Institute Potsdam, 
University of Basel, University of Cambridge, Case Western Reserve University, 
University of Chicago, Drexel University, Fermilab, the Institute for Advanced Study, 
the Japan Participation Group, Johns Hopkins University, the Joint Institute for 
Nuclear Astrophysics, the Kavli Institute for Particle Astrophysics and Cosmology, 
the Korean Scientist Group, the Chinese Academy of Sciences (LAMOST), 
Los Alamos National Laboratory, the Max-Planck-Institute for Astronomy (MPIA), 
the Max-Planck-Institute for Astrophysics (MPA), New Mexico State University, 
Ohio State University, University of Pittsburgh, University of Portsmouth, 
Princeton University, the United States Naval Observatory, 
and the University of Washington.

\appendix
\section{Query for Clean Photometry Stars}
Here we provide the query used to obtain the sample of isolated
and well measured stars used for the color-color diagrams. 
The query is to be run 
on the {\tt Stripe 82} database of the SDSS Catalog Arquive 
Server (CAS). \\

{\tt 
SELECT 

~ra, dec, run, camcol, field, 

~u, g, r, i, z, 

~psfMag\_u, psfMag\_g, psfMag\_r, psfMag\_i, psfMag\_z, flags,

~psfmagerr\_u, psfmagerr\_g, psfmagerr\_r, psfmagerr\_i, psfmagerr\_z

FROM 

~PhotoObjAll 

WHERE 

~~~~~((flags \& 0x10000000) != 0)

   ~AND ((flags \& 0x8100000c00a4) = 0)

   ~AND (((flags \& 0x400000000000) = 0) or 

~~~~~~(psfmagerr\_r <= 0.2 and psfmagerr\_i<= 0.2 and psfmagerr\_g<=0.2))

   ~AND (((flags \& 0x100000000000) = 0) or (flags \& 0x1000) = 0)

~AND (run = 106 or run = 206)

~AND type = 6

~AND mode = 1

~AND ra between 355 and 0
}

\bibliography{south}

\end{document}